\documentclass[lettersize,journal]{IEEEtran}

\usepackage{amsmath,amssymb,bm,mathtools}

\usepackage{graphicx}
\usepackage[caption=false,font=footnotesize,labelfont=rm,textfont=rm]{subfig}

\makeatletter
\renewcommand{\p@subfigure}{\thefigure}
\makeatother
\usepackage{stfloats}

\usepackage{array}

\usepackage{cite}
\usepackage{url}

\usepackage{xcolor}
\definecolor{revisionblue}{HTML}{0000FF}
\usepackage{color}
\usepackage{makecell}

\usepackage[ruled,linesnumbered]{algorithm2e}

\SetKwFor{ForE}{\textbf{for}}{\textbf{ do}}{\textbf{ end for}}
\SetKwFor{WhileE}{\textbf{while}}{\textbf{ do}}{\textbf{ end while}}
\SetKwIF{IfE}{ElseIfE}{ElseE}{\textbf{if}}{\textbf{ then}}{\textbf{ else if}}{\textbf{ else}}{\textbf{ end if}}

\SetAlFnt{\small}          
\SetAlCapFnt{\small}       
\SetAlCapNameFnt{\small}   
\SetAlgoNlRelativeSize{0}       
\SetInd{0.8em}{1.2em}           
\DontPrintSemicolon          

\usepackage{textcomp}
\usepackage{verbatim}
\usepackage{enumitem}
\RequirePackage{bibspacing}

\AtBeginDocument{
    \setlength{\jot}{1.1pt}
    \setlength{\abovedisplayskip}{1.1pt}
    \setlength{\abovedisplayshortskip}{1.1pt}
    \setlength{\belowdisplayskip}{1.1pt}
    \setlength{\belowdisplayshortskip}{1.1pt}
}
\definecolor{mygreen}{RGB}{252, 93, 3}   
\usepackage[colorlinks=true,
            linkcolor=mygreen,     
            anchorcolor=mygreen,
            citecolor=blue,      
            urlcolor=blue]{hyperref}

\usepackage{cleveref}
\crefname{algocf}{Algorithm}{Algorithms}
\Crefname{algocf}{Algorithm}{Algorithms}

\usepackage{xpatch}
\makeatletter
\ExplSyntaxOn
\cs_new:Npn \bibColoredItems #1#2
  { \clist_map_inline:nn {#2} { \cs_new:cpn {bib@colored@##1} {#1} } }
\ExplSyntaxOff
\newcommand\bib@setcolor[1]{%
  \ifcsname bib@colored@#1\endcsname
    \expanded{\noexpand\color{\csname bib@colored@#1\endcsname}}%
  \else
    \normalcolor
  \fi
}
\IfPackageLoadedTF{hyperref}{\@tempswatrue}{\@tempswafalse}
\if@tempswa
  \xpatchcmd\@bibitem {\H@item}{\bib@setcolor{#1}\H@item}{}{\PatchFailed}
  \xpatchcmd\@lbibitem{\H@item}{\bib@setcolor{#2}\H@item}{}{\PatchFailed}
\else
  \xpatchcmd\@bibitem {\item}  {\bib@setcolor{#1}\item}  {}{\PatchFailed}
  \xpatchcmd\@lbibitem{\item}  {\bib@setcolor{#2}\item}  {}{\PatchFailed}
\fi
\makeatother

\definecolor{revisionblue}{HTML}{0000FF}

\begin{document}

\title{Multipath Routing for Multi-Hop UAV Networks}

\author{Zhenyu Zhao,~\IEEEmembership{Graduate Student Member, IEEE}, Tiankui Zhang,~\IEEEmembership{Senior Member, IEEE}, Xiaoxia Xu,~\IEEEmembership{Member, IEEE}, Junjie Li, Yuanwei Liu,~\IEEEmembership{Fellow, IEEE}, Wenjuan Xing
\thanks{Zhenyu Zhao and Tiankui Zhang are with the School of Information and Communication Engineering, Beijing University of Posts and Telecommunications, Beijing 100876, China (e-mail: zhaozhenyu@bupt.edu.cn; zhangtiankui@bupt.edu.cn).}

\thanks{Xiaoxia Xu is with the School of Electronic Engineering and Computer Science, Queen Mary University of London, London E1 4NS, U.K. (e-mail: x.xiaoxia@qmul.ac.uk).}

\thanks{Junjie Li is with China Telecom Beijing Research Institute, Beijing 102200, China (e-mail: lijj28@chinatelecom.cn).}

\thanks{Yuanwei Liu  is with the Department of Electrical and Electronic Engineering, The University of Hong Kong, Hong Kong. (e-mail: yuanwei@hku.hk).}

\thanks{Wenjuan Xing  is with the School of Microelectronics and Communication Engineering, Chongqing University, Chongqing 401331, China. (e-mail: xingwj@chinatelecom.cn).}

}



\maketitle

\begin{abstract}

Multi-hop uncrewed aerial vehicle (UAV) networks are promising to extend the terrestrial network coverage. Existing multi-hop UAV networks employ a single routing path by selecting the next-hop forwarding node in a hop-by-hop manner, which leads to local congestion and increases traffic delays. In this paper, a novel traffic-adaptive multipath routing method is proposed for multi-hop UAV networks, which enables each UAV to dynamically split and forward traffic flows across multiple next-hop neighbors, thus meeting latency requirements of diverse traffic flows in dynamic mobile environments. An on-time packet delivery ratio maximization problem is formulated to determine the traffic splitting ratios at each hop. This sequential decision-making problem is modeled as a decentralized partially observable Markov decision process (Dec-POMDP). To solve this Dec-POMDP, a novel multi-agent deep reinforcement leaning (MADRL) algorithm, termed Independent Proximal Policy Optimization with Dirichlet Modeling (IPPO-DM), is developed. Specifically, the IPPO serves as the core optimization framework, where the Dirichlet distribution is leveraged to parameterize a continuous stochastic policy network on the probability simplex, inherently ensuring feasible traffic splitting ratios. Simulation results demonstrate that IPPO-DM outperforms benchmark schemes in terms of both delivery latency guarantee and packet loss performance.

\end{abstract}


\section{Introduction}
\IEEEPARstart{W}{ith} their high mobility, rapid deployability, and wide coverage capabilities, uncrewed aerial vehicles (UAVs) have been extensively deployed in scenarios such as post-disaster search and rescue, environmental monitoring, and aerial surveillance~\cite{UAV-intro-1, UAV-intro-2}. Depending on specific application requirements, UAVs can take on diverse roles within communication networks. On the one hand, UAVs may serve as aerial terminals, collecting various types of information and forwarding it to ground access points~\cite{UAV-sensing-1, UAV-sensing-2}. On the other hand, UAVs can operate as aerial base stations~\cite{UAV-BS-1, UAV-BS-2, UAV-BS-3} or computing nodes~\cite{UAV-MEC-1, UAV-MEC-2}, providing access services for ground users. In addition, within multi-UAV networks, they may serve as relay nodes, extending coverage through multi-hop transmission~\cite{UAV-relay-1, UAV-relay-2, UAV-relay-3}. \par

In typical multi-hop UAV networks, designing flexible and efficient routing strategies is critical to ensuring that data are reliably delivered to their intended destinations (e.g., ground nodes or a command center)~\cite{UAV-routing-des-BS-1, UAV-routing-des-BS-2}. However, frequent topology changes, unstable inter-UAV links, and constraints on storage and communication resources pose significant challenges for routing in such multi-hop UAV networks. Existing studies have devoted substantial effort to addressing these challenges, leading to a wide range of routing strategies tailored for multi-hop UAV networks~\cite{UAV-OLSR-2, UAV-AODV-routing-reactive, Ant-Colony-routing-hybrid,geo-routing,UAV-OLSR-1,q-learning-routing-1,q-learning-routing-2,q-learning-routing-3, swarm-learning-routing-1,swarm-learning-routing-2,trusted-learning-routing-1,trusted-learning-routing-2,anti-jamming-learning-routing,uav-satellite-learning-routing,coding-learning-routing}.\par

\subsection{Related Work}

In existing studies on multi-hop UAV network routing, design approaches can generally be categorized into protocol-based and learning-based schemes. The former relies on predefined rules to establish and maintain transmission paths, with advantages such as transparency and lightweight implementation overhead~\cite{UAV-OLSR-2, UAV-AODV-routing-reactive, Ant-Colony-routing-hybrid,geo-routing,UAV-OLSR-1}. The latter usually adopts reinforcement learning to dynamically optimize routing decisions, offering adaptability and flexibility in complex and rapidly changing network environments~\cite{q-learning-routing-1,q-learning-routing-2,q-learning-routing-3,swarm-learning-routing-1,swarm-learning-routing-2,trusted-learning-routing-1,trusted-learning-routing-2,anti-jamming-learning-routing,uav-satellite-learning-routing,coding-learning-routing}.

\subsubsection{Protocol-based routing approaches}

Based on~\cite{UAV-routing-survey}, protocol-based routing can be further classified into topology-based routing, geographic routing, and hybrid (combination of topology-based and geographic routing) routing. Topology-based routing primarily relies on network topology information and maintains routing tables through proactive, reactive, or hybrid mechanisms~\cite{UAV-OLSR-2, UAV-AODV-routing-reactive, Ant-Colony-routing-hybrid}. Geographic routing makes forwarding decisions based on the positional information of nodes~\cite{geo-routing}. Hybrid routing integrates the mechanisms of both topology-based and geographic routing~\cite{UAV-OLSR-1}.\par

In topology-based routing schemes, proactive routing maintains a global view of the network topology through periodic updates, thereby enabling rapid route selection, as exemplified by Optimized Link State Routing (OLSR). In~\cite{UAV-OLSR-2}, a neural network is incorporated into OLSR to jointly consider link stability, bandwidth, and energy consumption during path selection, with the goal of reducing latency and improving routing stability. Reactive routing, by contrast, establishes paths only when needed, such as in Ad hoc On-Demand Distance Vector (AODV). In~\cite{UAV-AODV-routing-reactive}, an order preference by similarity to ideal solution method is integrated into AODV to account for multiple factors, including hop count, stability, and residual energy, in the routing process. Hybrid routing combines both: proactive maintenance within local regions ensures availability, while reactive discovery handles long-distance communication. In~\cite{Ant-Colony-routing-hybrid}, a distributed hybrid ant colony routing protocol is proposed, which integrates proactive pheromone maintenance with reactive on-demand ant searching, and incorporates link quality prediction to achieve a trade-off between low latency and low overhead in highly dynamic UAV networks.\par

Geographic routing utilizes the positional information of UAVs to make forwarding decisions. Such schemes typically adopt a greedy forwarding strategy, where packets are forwarded to the neighbor closest to the destination. In~\cite{geo-routing}, a three-dimensional geographic routing protocol based on the concept of effective transmission range is proposed. This protocol improves the greedy forwarding strategy by selecting the neighbor with the highest link utility and applies a 3D perimeter recovery mechanism when candidate nodes are insufficient.\par

Hybrid routing combines topology-based and geographic mechanisms, for example by incorporating positional information into topology maintenance. In~\cite{UAV-OLSR-1}, an enhanced OLSR protocol with location awareness is proposed, where factors such as node position, velocity, and direction are integrated into the multipoint relay selection process, significantly improving routing stability.

\subsubsection{Learning-based routing approaches}
Learning-based routing schemes typically employ reinforcement learning and other intelligent approaches to acquire effective routing strategies. Existing studies have explored multiple directions, such as optimizing hop-by-hop routing, enhancing coordination in large-scale UAV swarms, and strengthening communication security~\cite{q-learning-routing-1,q-learning-routing-2,q-learning-routing-3,swarm-learning-routing-1,swarm-learning-routing-2,trusted-learning-routing-1,trusted-learning-routing-2,anti-jamming-learning-routing,uav-satellite-learning-routing,coding-learning-routing}.\par

For hop-by-hop routing optimization, the goal is to learn effective routing strategies that leverage information from neighboring UAV nodes (e.g., residual energy, link quality) to make effective next-hop forwarding decisions. Studies such as~\cite{q-learning-routing-1,q-learning-routing-2,q-learning-routing-3} design Q-learning-based reward functions that exploit neighbor information to select the most suitable next-hop node, thereby enabling routing strategies that reduce end-to-end delay, improve delivery ratio, and balance energy consumption.\par

To enhance communications in UAV swarm scenarios,~\cite{swarm-learning-routing-1} proposed a method that combines multi-agent reinforcement learning (MARL) with adaptive communication mechanisms, which reduces flooding overhead and improves routing coordination. Similarly,~\cite{swarm-learning-routing-2} introduced an intelligent cluster-based routing scheme that addresses routing instability and load imbalance through multi-factor weight optimization and adaptive cluster-head selection.\par

In scenarios with high security requirements,~\cite{trusted-learning-routing-1} proposes a trust-aware routing framework that integrates blockchain with MARL. By quantifying and disseminating node trust, it mitigates malicious behavior and improves network resilience. Another study,~\cite{trusted-learning-routing-2} develops a reinforcement-learning-based recovery scheme that leverages node-importance modeling and learning-driven path reconstruction to rapidly restore connectivity under targeted routing attacks.\par

Beyond the above directions, learning-based routing has also been explored in areas such as anti-jamming, space–air integration, and coding-aware routing. For instance,~\cite{anti-jamming-learning-routing} proposed a reinforcement-learning-driven scheme that jointly optimizes energy efficiency, anti-jamming capability, and quality of service in multi-hop UAV communications. In UAV–satellite integrated internet of things scenarios,~\cite{uav-satellite-learning-routing} developed a two-layer deep reinforcement learning framework that ensures reliable long-distance transmission and low-latency local links through hierarchical decision-making and quality of service (QoS)-aware rewards. Additionally,~\cite{coding-learning-routing} designed a coding-aware routing protocol that integrates deep reinforcement learning with network coding opportunity recognition to improve throughput while reducing redundancy.

\subsection {Motivations and Challenges}

\subsubsection{Motivations}
Current studies on multi-hop UAV network routing have made progress in optimizing generic metrics such as energy efficiency and average latency. However, existing schemes face significant limitations in handling complex mission scenarios, which motivates us to develop multipath-enabled and traffic-adaptive routing schemes. The specific motivations are discussed as follows:

\begin{itemize}
\item \textit{From Single-path to Multipath:} 
UAVs are typically constrained by limited on-board buffer size and communication bandwidth. In conventional single-path routing, the entire traffic flow is directed to a single next-hop neighbor. When handling large-scale data, this paradigm tends to overload specific relay nodes, rapidly exhausting their buffer capacity. This leads to congestion-induced packet loss and high queuing delays, creating distinct bottlenecks in the network. To address this, multipath forwarding is essential. By splitting and forwarding traffic flows to multiple next-hop neighbors in parallel, the network can effectively balance the load, alleviate bottlenecks on individual links, and enhance the overall packet delivery ratio.

\item \textit{Traffic Heterogeneity Awareness:} 
Traffic flows carried by UAV networks exhibit significant heterogeneity in terms of delay sensitivity and data volume. For instance, traffic generated by target recognition tasks is typically highly latency-sensitive but data-light~\cite{task-recognition-task}, whereas data analysis tasks involve large-scale datasets with looser delay constraints~\cite{data-analysis-task}. However, existing routing schemes often overlook these characteristics and treat diverse flows homogeneously. This uniform approach reduces performance, resulting in high latency for critical missions or congestion caused by large data flows. Therefore, it is necessary to design routing mechanisms that consider these characteristics to provide differentiated forwarding services.
\end{itemize}

\subsubsection{Challenges}
Designing a traffic-adaptive multipath routing scheme for multi-hop UAV networks is significantly more complex compared to terrestrial networks. We summarize the key challenges as follows:

\begin{itemize}
\item \textit{Highly Dynamic Topology and Volatile Links:} 
Unlike stable terrestrial environments, UAV networks are characterized by high mobility. The network topology can evolve within seconds, and link quality fluctuates in milliseconds. This poses a severe challenge for routing stability, as the set of feasible paths changes continuously. Strategies derived from historical network states often become invalid rapidly, requiring the routing method to adapt to environmental changes in real-time.

\item \textit{Decentralized Decision-Making under Partial Observability:} 
UAV swarms typically operate without a centralized controller for global path planning. Each UAV node must make routing decisions based solely on partial observability (e.g., local queue status and one-hop neighbor information) without access to downstream congestion status or global end-to-end delay. It is challenging to guarantee long-term QoS requirements for heterogeneous flows based only on such limited local views.

\item \textit{Complexity of Joint Splitting and Forwarding:} 
Implementing traffic-adaptive multipath routing expands the decision space significantly. The system must not only select a set of candidate neighbors but also determine the optimal traffic splitting ratios among them. This involves a complex trade-off, confining latency-sensitive flows to optimal paths for speed, while dispersing large-volume flows across multiple neighbors to prevent congestion. Balancing these conflicting objectives under the aforementioned dynamic and decentralized constraints constitutes a complex sequential decision-making dilemma.
\end{itemize}

\subsection {Contributions}

To address the aforementioned limitations, this paper proposes a novel traffic-adaptive multipath routing method tailored for multi-hop UAV networks. By explicitly perceiving traffic heterogeneity in terms of data volume and delay sensitivity, the proposed method enables UAV nodes to make distributed and adaptive traffic splitting decisions over aggregated traffic, thereby simultaneously harmonizing traffic-specific latency satisfaction and network-wide load balancing. Specifically, we mathematically characterize the adaptive traffic splitting process as an optimization problem aimed at maximizing the on-time packet delivery ratio. This problem is cast as a Decentralized Partially Observable Markov Decision Process (Dec-POMDP) and solved via a novel algorithm named Independent Proximal Policy Optimization with Dirichlet Modeling (IPPO-DM). This method integrates IPPO with a Dirichlet policy head to directly learn continuous and feasible splitting ratios on the probability simplex.\par

To the best of our knowledge, this is the first study to propose a traffic-adaptive multipath routing algorithm specifically tailored for multi-hop UAV networks. The main contributions of this paper are summarized as follows:

\begin{itemize}
\item We propose a traffic-adaptive multipath routing framework for multi-hop UAV networks, which enables each UAV to dynamically split and forward traffic flows across multiple next-hop neighbors, thus mitigating traffic congestion in conventional single-path routing schemes. We first employ a priority-based sub-queue mechanism to classify traffic according to delay sensitivity, and then perform an adaptive traffic splitting strategy that forwards data from non-empty high-priority queues to multiple neighboring nodes in parallel. We formulate the multipath routing optimization problem, which sequentially determines the traffic splitting ratios at each hop. The key objective is to maximize the on-time packet delivery ratio.

\item We recast the sequential multipath routing optimization problem as a Dec-POMDP. To solve this Dec-POMDP, we develop a MADRL algorithm named IPPO-DM. Specifically, we adopt the IPPO framework to enable robust distributed policy learning under partial observability, while incorporating a Dirichlet-parameterized policy to directly learn continuous traffic splitting actions on the probability simplex, thus inherently satisfying validity constraints. Furthermore, we introduce a traffic-aware resampling mechanism during the execution phase. This mechanism adaptively adjusts the number of forwarding paths based on real-time queue states, enhancing routing flexibility and efficiency without altering the learned policy structure.

\item We evaluate the proposed scheme under various traffic loads, network scales, and forwarding-node constraints. We consider two representative baseline routing algorithms, i.e., a heuristic average traffic splitting algorithm, and a conventional single-path greedy routing algorithm. Experimental results demonstrate that IPPO-DM consistently outperforms these baselines, achieving higher on-time packet delivery ratios and lower packet loss.

\end{itemize}

The structure of the remaining part of this paper is as follows. Section II introduces the system model and problem formulation. Section III demonstrates the problem solution. Section IV presents the numerical results. Finally, conclusions are summarized in Section V.\par

\begin{figure}
    \centering
    \includegraphics[width=1\linewidth]{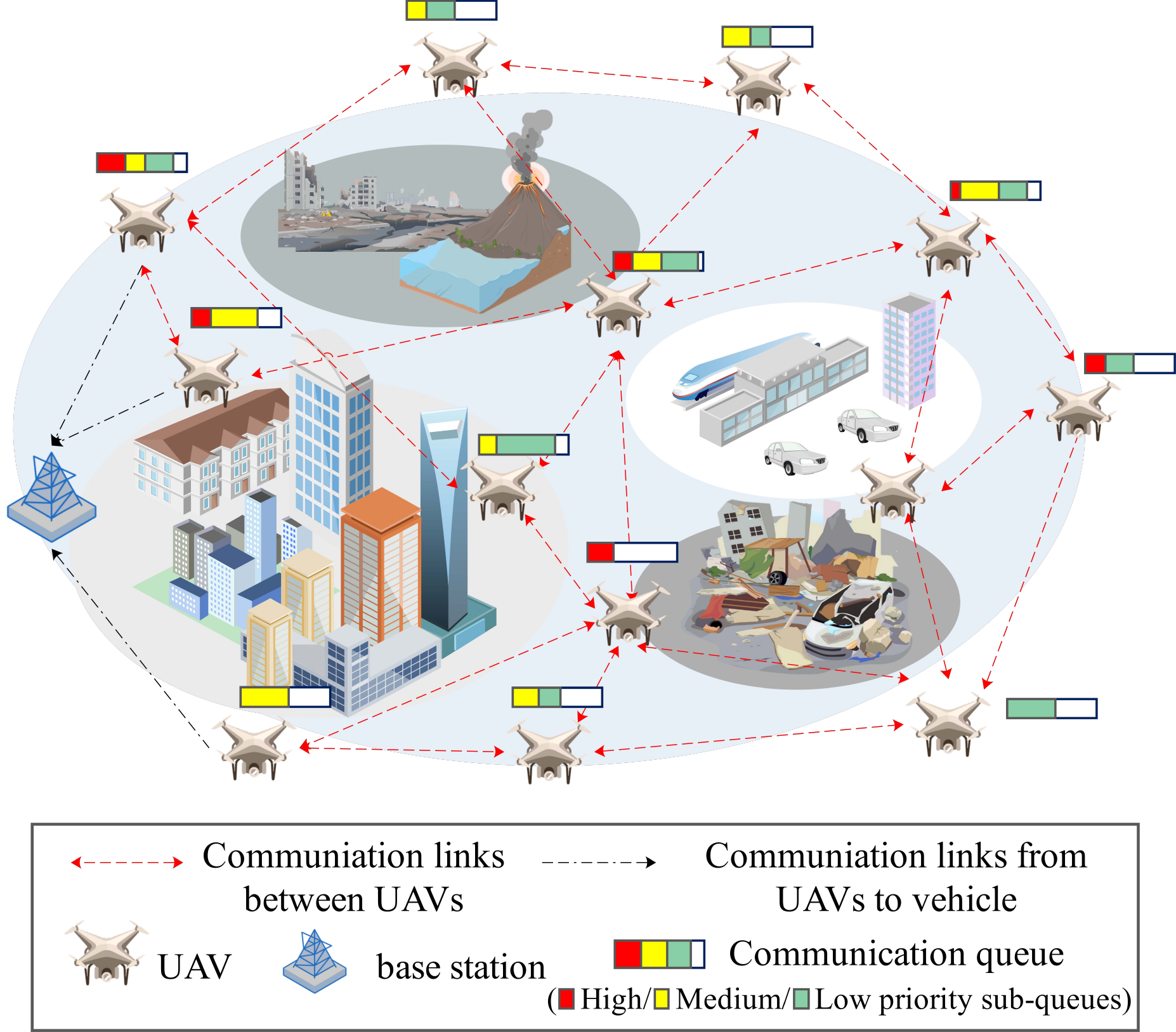}
    \caption{Multi-UAV and Ground Base Station Collaborative System.}
    \label{system-model}
\end{figure}

\section{ System Model And Problem Formulation}
\subsection{System model}

As shown in Fig.~\ref{system-model}, we consider a network architecture where multiple UAVs cooperate with a ground base station (GBS). The UAVs perform sensing or inspection tasks and transmit the collected data to the GBS for processing and analysis.\par

The set of UAVs is denoted by ${{\cal M}} \buildrel \Delta \over = {\rm{\{1,}}...{\rm{,}}M{\rm{\}}}$, and the GBS is denoted by $k$. Let $T$ denote the total flight duration of UAVs. Within $T$, each UAV follows its designated trajectory. The flight duration is discretized into $T_{\rm max}$ time slots of equal length $\Delta t = T/T_{\rm max}$, represented as ${\cal T} = \{1, \dots, T_{\rm max}\}$. The time slot length is chosen sufficiently small so that UAV positions can be regarded as static within each slot, i.e.,
\begin{equation}\label{time-slot-constraint}
\Delta t \le \Delta {t^{\max }}.
\end{equation}The position of UAV $m$ at time slot $t$ is denoted by ${{\bm l}_m}(t) = [x_m^{\rm uav}(t), y_m^{\rm uav}(t), z_m^{\rm uav}(t)]$, while the GBS $k$ is fixed at ${{\bm l}_k} = [x_k^{\rm veh}, y_k^{\rm veh}, 0]$.\par

In each time slot $t$, a UAV generates task-related traffic with probability ${p^{\rm traf}}$. Let $U$ denote the total number of traffic flows generated by all UAVs during period $T$, and define the traffic set as ${{\cal U}} = \{1, \dots, U\}$. For each traffic $u \in {\cal U}$, its deadline and data volume are denoted by $t_u^{\rm end}$ and $I_u$, respectively. Without loss of generality, each traffic flow is segmented into packets of size $I_{\rm pkt}$. If the last segment is smaller than $I_{\rm pkt}$, it is padded with default symbols to maintain consistency. The number of packets for traffic $u$ is ${W_u}$, with the corresponding packet set as ${{\cal W}_u} = \{1, \dots, W_u\}$. Each packet $w \in {\cal W}_u$ inherits the deadline of its parent traffic, i.e., $t_w^{\rm end} = t_u^{\rm end}$. The set of all packets generated during $T$ is denoted by ${{\cal W}} = \bigcup_{u \in {\cal U}} {\cal W}_u$.

\emph{1)  Communication Model:} Since UAVs operate at altitude with largely unobstructed links, line-of-sight (LoS) channels dominate UAV-to-UAV communications. At time slot $t$, the distance between UAV $m$ and UAV $m'$ is given by ${d_{m,m'}}(t) = \|{{\bm l}_m}(t) - {{\bm l}_{m'}}(t)\|_2$. Let ${\beta _0}$ denote the channel gain at a reference distance of 1 meter, then the channel gain between UAVs $m$ and $m'$ is expressed as ${h_{m,m'}}(t) = {\beta _0}/d_{m,m'}^{2}(t)$.\par

Each UAV can forward data to at most $N+1$ neighboring nodes (other UAVs or the GBS) in each time slot. Its transmit power is evenly allocated among these nodes, such that the transmit power from UAV $m$ to any neighbor is ${p_m} = p_m^{\max }/(N+1)$, where $p_m^{\max }$ is the maximum transmit power of UAV $m$.\par

When UAV $m$ transmits to UAV $m'$ at time slot $t$, the set of co-frequency nodes within $m$’s communication range is denoted by ${\phi _{m'}}(t)$. The interference at UAV $m'$ is then given by $I_{m,m'}(t) = \sum\nolimits_{i \in \phi_{m'}(t), i \ne m} p_i h_{i,m'}(t)$. Let ${N_0}$ denote the noise spectral density. The system employs $B$ orthogonal sub-channels, each of bandwidth $b$. Thus, the signal-to-interference-plus-noise ratio (SINR) between UAV $m$ and UAV $m'$ at time slot $t$ is $\vartheta_{m,m'}(t) = \frac{p_m h_{m,m'}(t)}{bN_0 + I_{m,m'}(t)}$ and the corresponding transmission rate is
\begin{align}
R_{m,m'}(t) = &b \log_2 \big(1 + \vartheta_{m,m'}(t)\big), \nonumber \\
&\forall t, m, m' \in {\cal M}, m \ne m'.
\label{comm-UAVs}
\end{align}Finally, the number of packets that UAV $m$ can deliver to UAV $m'$ within time slot $t$ is $g_{m,m'}(t) = \frac{R_{m,m'}(t)\Delta t}{I_{\rm pkt}}.$\par

Unlike UAV-to-UAV links, ground environments typically contain obstacles, and thus UAV-to-GBS channels must account for both LoS and Non-Line-of-Sight (NLoS) conditions. At time slot $t$, the distance and elevation angle between GBS $k$ and UAV $m$ are denoted by $d_{m,k}(t) = \| {\bm l}_m(t) - {\bm l}_k \|_2$, $ 
\theta_{m,k}(t) = \frac{180^\circ}{\pi} \arcsin \left( \frac{z_m^{\rm uav}(t)}{d_{m,k}(t)} \right),$ respectively. The probability of an LoS link is modeled as $P_{m,k}^{\rm LoS}(t) = \frac{1}{1 + D_1 \exp \left( - D_2 \left[ \theta_{m,k}(t) - D_1 \right] \right)},$ where $D_1$ and $D_2$ are environment-dependent parameters. Accordingly, the NLoS probability is $P_{m,k}^{\rm NLoS}(t) = 1 - P_{m,k}^{\rm LoS}(t)$. Following~\cite{LosNLos-channel-model}, the path loss is given by $L_{m,k}^\xi(t) = \eta_\xi \left( \frac{4 \pi f_c d_{m,k}(t)}{D_3} \right)^\hbar$, $ \xi \in \{\rm LoS, \rm NLoS\},$ where $\hbar$ is the path-loss exponent, $\eta_\xi$ is the excessive path loss factor, $f_c$ is the carrier frequency, and $D_3$ is the speed of light. Based on both LoS and NLoS conditions, the expected channel gain is
\begin{align}
h_{m,k}(t) = &\frac{1}{P_{m,k}^{\rm LoS}(t) L_{m,k}^{\rm LoS}(t) + P_{m,k}^{\rm NLoS}(t) L_{m,k}^{\rm NLoS}(t)},  \nonumber \\
&\forall m,k,t.
\label{gain-UAV-GBS}
\end{align}

When UAV $m$ transmits to GBS $k$ at time slot $t$, the interference at the GBS from other UAVs is $I_{m,k}(t) = \sum_{i \in \phi_k(t), i \ne m} p_i h_{i,k}(t),$ where $\phi_k(t)$ denotes the set of co-frequency nodes within GBS $k$’s communication range. Consequently, the SINR of the link between UAV $m$ and GBS $k$ is $\vartheta_{m,k}(t) = \frac{p_m h_{m,k}(t)}{b N_0 + I_{m,k}(t)},$ and the corresponding transmission rate is
\begin{equation}
R_{m,k}(t) = b \log_2 \left(1 + \vartheta_{m,k}(t)\right), \quad \forall m,k,t.
\label{comm-uav-GBS}
\end{equation}
Then, the number of packets UAV $m$ can transmit to GBS $k$ during time slot $t$ is $g_{m,k}(t) = \frac{R_{m,k}(t) \Delta t}{I_{\rm pkt}}.$\par

To guarantee reliable data transmission, let $\vartheta^{\min}$ denote the minimum SINR threshold required for establishing a communication link between two nodes~\cite{SINR-threshold}. Accordingly, at time slot $t$, UAV $m$ can communicate with GBS $k$ only if $\vartheta_{m,k}(t) \geq \vartheta^{\min}$. Similarly, the set of UAV neighbors that are reachable from UAV $m$ at time slot $t$ is defined as $\psi_m(t)$, and must satisfy
\begin{equation}
\vartheta_{m,m'}(t) \geq \vartheta^{\min}, \quad \forall m' \in \psi_m(t).
\label{SINR-thre}
\end{equation}
Furthermore, UAV $m$ may randomly select up to $N$ UAVs from $\psi_m(t)$ as candidate next-hop relays, denoted by $\psi_m^{\rm com}(t) \subseteq \psi_m(t)$, with the constraint $|\psi_m^{\rm com}(t)| \leq N$.

\emph{2) Queueing Model:} To prioritize latency-sensitive traffic, the communication queue of each UAV is divided into three sub-queues corresponding to high, medium, and low priority levels, denoted as $Q_m(t) = \{ Q_m^1(t), Q_m^2(t), Q_m^3(t)\},$ where $Q_m^1(t)$, $Q_m^2(t)$, and $Q_m^3(t)$ represent the high-, medium-, and low-priority sub-queues, respectively. At each time slot $t$, the active transmission queue of UAV $m$ is defined as the highest-priority non-empty sub-queue, denoted by $Q_m^{\rm sel}(t)$. Its priority index is $\pi_m^{\rm sel}(t) \in \{1,2,3\}$, and the number of packets awaiting forwarding in this queue is $q_m^{\rm sel}(t)$.  \par

Let $q_m^{\max}$ denote the maximum buffer capacity (in packets) of UAV $m$’s communication queue $Q_m(t)$, and let $q_m^{\rm free}(t)$ denote the available buffer capacity at time slot $t$. The number of packets in the three sub-queues are denoted by $q_m^1(t)$, $q_m^2(t)$, and $q_m^3(t)$, respectively. These variables satisfy
\begin{equation}
    q_m^1(t) + q_m^2(t) + q_m^3(t) + q_m^{\rm free}(t) = q_m^{\max},  \forall m,t.
    \label{queue-capa-constr}
\end{equation}

Incoming or locally generated traffic is placed into different sub-queues according to their deadlines. Specifically, at time slot $t$, the priority of traffic $u \in \mathcal{U}$ is determined by its urgency, denoted as $u_{\rm prio}(t)$, and is given by
\begin{equation}
u_{\rm prio}(t) = 
\begin{cases} 
1, & \text{if } t_u^{\rm end} - t \leq 0.5 \, \text{s}, \\ 
2, & \text{if } 0.5 \, \text{s} < t_u^{\rm end} - t \leq 1 \, \text{s}, \\ 
3, & \text{if } t_u^{\rm end} - t > 1 \, \text{s}.
\end{cases}
\label{traffic_prio}
\end{equation}

\emph{3) Forwarding Model:} At time slot $t$, if UAV $m$ can communicate with GBS $k$, it forwards as many packets as possible from $Q_m^{\rm sel}(t)$ to the GBS. The number of successfully transmitted packets is denoted by $q_{m,k}^{\rm trans}(t)$, subject to $q_{m , k}^{\rm trans}(t) \le g_{m,k}(t), \forall m, t.$ Packets that cannot be transmitted remain in the queue for future transmission.  \par

If UAV $m$ cannot directly communicate with the GBS, it distributes the packets in $Q_m^{\rm sel}(t)$ among its neighbors. The routing decision at time slot $t$ is defined as ${\bm a}_m(t) = \big[\,a_{m , m'}(t)\,\big]_{m' \in \{0\} \cup \psi_m^{\rm com}(t)},$ where $a_{m , 0}(t)$ denotes the fraction of packets retained locally, and $a_{m , m'}(t), \forall m' \in \psi_m^{\rm com}(t)$ denotes the fraction allocated to neighbor UAV $m'$. These variables satisfy
\begin{align} &0 \le a_{m , m'}(t) \le 1, \nonumber \\ &\forall m \in {\cal M}, \, m' \in \{0\} \cup \psi_m^{\text{com}}(t), \, t \in {\cal T}, \label{forward-0-1} \\ &\sum_{\mathclap{m' \in \{0\} \cup \psi_m^{\text{com}}(t)}} a_{m , m'}(t) = 1, \quad \forall t \in \mathcal{T}. \label{forward-sum} \end{align}Accordingly, UAV $m$ partitions the packets in $Q_m^{\rm sel}(t)$ into non-overlapping segments based on its routing decision ${\bm a}_m(t)$.\par

The number of packets actually transmitted to neighbor $m'$ is constrained by the link capacity, the neighbor’s available buffer, and the forwarding decision, i.e.,
\begin{align}
q_{m , m'}^{\rm trans}(t) \le \min (& g_{m,m'}(t), \, q_{m'}^{\rm free}(t), \nonumber\\
& a_{m , m'}(t) \cdot q_m^{\rm sel}(t) ).
\label{UAVs-trans-packs}
\end{align}

Similarly, define the receive set of UAV $m$ at slot $t$ as $\psi^{\rm in}_m(t) = 
\{\, m' \in \mathcal{M} \setminus \{m\} \mid m \in \psi_{m'}^{\rm com}(t)\}$, which denotes the set of UAVs that may forward packets to $m$. Accordingly, the total number of packets that UAV $m$ can receive at slot $t$ 
is constrained by its available buffer capacity,
\begin{equation}
\sum_{m' \in \psi^{\rm in}_m(t)} q_{m' , m}^{\rm trans}(t) 
\;\le\; q_m^{\rm free}(t), 
\quad \forall m \in \mathcal{M}, \; t \in \mathcal{T}.
\label{UAVs-recei-packs}
\end{equation}

If $q_{m , m'}^{\rm trans}(t) < a_{m , m'}(t) \cdot q_m^{\rm sel}(t)$, the excess packets are dropped. The number of discarded packets is $q_m^{\rm loss}(t) = a_{m , m'}(t) \cdot q_m^{\rm sel}(t) - q_{m , m'}^{\rm trans}(t).$ The overall packet loss ratio during $T$ is then expressed as
\begin{equation}
\varphi_{\rm loss} = \frac{\sum_{t=1}^T \sum_{m=1}^M q_m^{\rm loss}(t)}{\sum_{u=1}^U W_u},
\label{loss-ratio}
\end{equation}
which must satisfy
\begin{equation}
\varphi_{\rm loss} \le \varphi_{\rm loss}^{\max}.
\label{loss-ratio-thre}
\end{equation}

Let $m_w^0$ denote the source UAV of packet $w \in {\cal W}$, and let $\{ m_w^1, \ldots, m_w^d \}$ denote its relay sequence. The routing path of packet $w$ is represented as $s_w = \{ m_w^0, m_w^1, \ldots, m_w^d, k \}$. The corresponding arrival time is $t_w^{\rm arr} = (d+1)\Delta t.$ Define $\kappa_{\rm pack} = \{ w \in {\cal W} \mid t_w^{\rm arr} \le t_w^{\rm end} \}$ as the set of packets that arrive no later than their deadlines. The on-time delivery ratio is given by
\begin{equation}
\eta_{\rm pack} = \frac{|\kappa_{\rm pack}|}{\sum_{u=1}^U W_u},
\label{pack-deli-ratio}
\end{equation}where $|\kappa_{\rm pack}|$ denotes the number of packets delivered within deadlines during $T$.

\subsection{Problem Formulation}
This work aims to develop a distributed traffic-adaptive multipath routing strategy that maximizes the on-time delivery ratio of traffic. Let ${\bf A} = \{ {\bm a}_m(t) \mid \forall t \in {\cal T}, \forall m \in {\cal M}\}$ denote the set of routing decision variables for all UAVs. The optimization problem can be formulated as

\begin{subequations} \label{P1}
\begin{align} 
{\rm{(P1):}} \quad & \mathop{\max}\limits_{\bf A} \ \eta_{\mathrm{pack}} \nonumber \\ 
{\rm{s.t.}}\quad & \eqref{time-slot-constraint},\ \eqref{SINR-thre},\ \eqref{forward-0-1}-\eqref{UAVs-recei-packs},\ \eqref{loss-ratio-thre}. \nonumber
\end{align}
\end{subequations}Here, \eqref{time-slot-constraint} specifies the maximum slot duration, while \eqref{SINR-thre} enforces the minimum SINR required for communication between nodes. Constraints \eqref{forward-0-1} and \eqref{forward-sum} ensure that each routing decision component lies within $[0,1]$ and that their sum equals one. Constraint \eqref{UAVs-trans-packs} limits the packets a UAV can forward to a neighbor, ensuring it does not exceed either the link capacity or the neighbor’s buffer size, whereas \eqref{UAVs-recei-packs} restricts the packets a UAV can receive so that they do not exceed its available queue capacity. Finally, \eqref{loss-ratio-thre} imposes an upper bound on the overall packet loss ratio.

\section{Problem Solution}
In dynamic multi-hop UAV networks, making routing decisions in a distributed environment is inherently complex. Moreover, routing strategies must account for the data volume and delay requirements of heterogeneous traffic and perform adaptive traffic splitting and forwarding accordingly, which further increases the difficulty of the problem. To address these challenges, we reformulate  (\hyperref[P1]{P1}) as a Dec-POMDP, enabling agents to adaptively adjust routing strategies through interactive learning with the dynamic environment. To obtain efficient routing policies, we propose the IPPO-DM algorithm. In this framework, IPPO is employed to achieve robust policy learning in distributed settings, where the policy network parameterizes a Dirichlet distribution to inherently model continuous traffic splitting actions on the probability simplex.

\subsection{Dec-POMDP Formulation}
We formally model the distributed routing problem (\hyperref[P1]{P1}) as a Dec-POMDP, which is solved within the MARL paradigm~\cite{MARL}. In this formulation, each UAV is modeled as an independent agent that selects traffic splitting actions at every time slot based solely on its local observations. After executing an action, the agent receives an immediate reward from the environment and transitions to the next state.\par

\emph{1) Observation Space:} At time slot $t$, the local observation of UAV $m$ is defined as ${\bm o}_m(t) = \big[{\bm o}_m^{\text{own}}(t), \, {\bm o}_m^{\text{neigh}}(t)\big],$ where ${\bm o}_m^{\text{own}}(t)$ denotes the self-state vector of UAV $m$, and ${\bm o}_m^{\text{neigh}}(t)$ denotes the state vector of its neighboring UAVs. The self-state vector ${\bm o}_m^{\text{own}}(t)$ consists of the distance $d_{m,k}(t)$ between UAV $m$ and the GBS $k$, the priority index $\pi_m^{\text{sel}}(t)$ of the selected sub-queue $Q_m^{\text{sel}}(t)$, and the number of packets waiting for transmission in that sub-queue $q_m^{\text{sel}}(t)$. Formally,
\begin{equation}
    {\bm o}_m^{\text{own}}(t) = [d_{m,k}(t), \, \pi_m^{\text{sel}}(t), \, q_m^{\text{sel}}(t)].
    \label{self-state}
\end{equation}

Since each UAV can maintain at most $N$ candidate next-hop relays, the neighbor-state vector ${\bm o}_m^{\text{neigh}}(t)$ is constructed as a fixed-length vector containing the states of up to $N$ neighbors. For consistency, neighbors in $\psi_m^{\text{com}}(t)$ are sorted in ascending order of their global identifiers, forming a sequence $s_m(t)$ of length $|s_m(t)| \le N$. The global identifier of the $n$-th neighbor is denoted as $m' = s_m^{(n)}(t)$. Following this order, the state vector of each neighbor is inserted into ${\bm o}_m^{\text{neigh}}(t)$, which can be written as
\begin{equation}
    {\bm o}_m^{\text{neigh}}(t) = \big[{\bm o}_{m,1}^{\text{neigh}}(t), \ldots, {\bm o}_{m,N}^{\text{neigh}}(t)\big].
    \label{neighbor-state}
\end{equation}If $|s_m(t)| < N$, the remaining positions are padded with zeros.\par

The $n$-th neighbor-state vector ${\bm o}_{m,n}^{\text{neigh}}(t)$ includes the neighbor’s identifier $m'$, its distance to the GBS $d_{m',k}(t)$, and a metric $c_{m,m'}(t)$ that reflects the joint impact of the receiving UAV’s available buffer and the link capacity. Formally, the $n$-th neighbor-state vector ${\bm o}_{m,n}^{\text{neigh}}(t)$ is given by 
\begin{equation}
    {\bm o}_{m,n}^{\text{neigh}}(t) = 
    \begin{cases} 
    [m', \, d_{m',k}(t), \, c_{m,m'}(t)], & \text{if } n \leq |\psi_m^{\text{com}}(t)|, \\[6pt] 
    [0, 0, 0], & \text{otherwise.} 
    \end{cases}
    \label{each-negihbor-state}
\end{equation}

In addition, to improve training stability, all features in the observation vector are linearly normalized according to their respective ranges.

\emph{2) Action Space:} At time slot $t$, UAV $m$ forwards packets from its sub-queue $Q_m^{\text{sel}}(t)$ either to its neighbors or retains them locally. This decision is represented by the action vector
\begin{equation}
    {\bm a}_m(t) = [a_{m,0}(t), \, a_{m,1}(t), \ldots, a_{m,N}(t)],  \forall m,t,
    \label{action}
\end{equation}where $a_{m,0}(t)$ represents the proportion of packets retained locally, and $a_{m,n}(t)$ represents the proportion forwarded to the $n$-th neighbor in $s_m(t)$. These variables satisfy $0 \leq a_{m,n}(t) \leq 1$ for all $n \in \{0,1,\ldots,N\}$. The action vector must satisfy the normalization constraint
\begin{equation}
    \sum_{n=0}^{N} a_{m,n}(t) = 1, \quad \forall m,t.
    \label{action-sum}
\end{equation}

\emph{3) Reward Function:} To evaluate the impact of each traffic splitting decision, we define the forwarding reward associated with allocation proportion $a_{m,n}(t)$ as $r_{m,n}(t)$. The overall routing reward is then obtained as the weighted sum of all forwarding rewards,
\begin{equation}
    r_m(t) = \sum_{n=0}^N a_{m,n}(t) \cdot r_{m,n}(t), \quad \forall m,t.
    \label{reward-fun}
\end{equation}

According to (\hyperref[P1]{P1}), each UAV must balance two objectives when making routing decisions: (i) maximizing the on-time delivery ratio of packets and (ii) reducing packet loss due to limited link capacity or buffer overflow. To capture both aspects, the forwarding reward $r_{m,n}(t)$ is defined as a weighted combination of the path forwarding reward $r_{m,n}^{\text{st}}(t)$, which reflects the impact of path length on timely delivery, and the loss-tolerance reward $r_{m,n}^{\text{tol}}(t)$, which measures the risk of packet loss,
\begin{equation}
    r_{m,n}(t) = w_m(t) \cdot r_{m,n}^{\text{st}}(t) + (1 - w_m(t)) \cdot r_{m,n}^{\text{tol}}(t),  \forall m,t.
    \label{component-reward}
\end{equation}

Furthermore, the weight $w_m(t)$ is linearly mapped to the queue length of the sub-queue $q_m^{\text{sel}}(t)$. When the queue is lightly loaded, $w_m(t)$ is large, prioritizing shorter routing paths. As the queue grows, $w_m(t)$ decreases, placing greater emphasis on avoiding packet loss. Formally,
\begin{equation}
    w_m(t) = k \cdot q_m^{\text{sel}}(t) + b, \quad \forall m,t,
    \label{reward-weight-para}
\end{equation}where $k = (w_{\min} - w_{\max})/(q_{\max} - q_{\min})$ and $b = w_{\max} - k \cdot q_{\min}$. Here, $w_{\min}$ and $w_{\max}$ are the extreme values of the weight parameter, and $q_{\min}, q_{\max}$ denote the minimum and maximum occupancy of the selected sub-queue.\par

With the weight determined, the two reward components are defined as follows.
\begin{equation}
r_{m,n}^{\text{st}}(t) = 
\begin{cases} 
d_{m,k}(t) - d_{m',k}(t), & 
\begin{aligned}[t]
&\text{if } d_{m,k}(t) > d_{m',k}(t), \\
& n \neq 0,
\end{aligned} \\[6pt]
- r^{\text{st}}(\pi_m^{\text{sel}}(t)), & \text{otherwise.}
\end{cases}
\end{equation}where the reward is positive if the next-hop neighbor reduces the distance to the GBS, and negative if no progress is made. The penalty is scaled according to the urgency level $\pi_m^{\text{sel}}(t)$, i.e., more urgent traffic incurs a higher penalty for poor forwarding decisions.\par

The loss-tolerance reward reflects the extent to which the planned forwarding volume aligns with the underlying transmission and buffer constraints. Specifically, if no forwarding is chosen ($n=0$), a fixed penalty $-r_0^{\text{tol}}$ is imposed. Otherwise, the reward is obtained through a smooth mapping that captures the discrepancy between the intended forwarding volume and an indicator of feasible transmission capability, given by
\begin{equation}
r_{m,n}^{\text{tol}}(t) =
\begin{cases}
- r_0^{\text{tol}}, & \text{if } n = 0, \\[6pt]
\begin{aligned}[t]
& \tanh \Big( k_1 \big(a_{m,n}(t) \cdot q_m^{\text{sel}}(t) \\
& \qquad - c_{m,m'}(t)\big) + k_2 \Big),
\end{aligned} 
& \text{if } n \neq 0,
\end{cases}
\label{loss-tolerance-reward}
\end{equation}where $c_{m,m'}(t)$ is the indicator of feasible transmission capability, and $k_1, k_2$ are tunable parameters controlling the slope and center of the reward curve.

\subsection{The Proposed IPPO-DM Algorithm}

Building on the Dec-POMDP formulation, we now detail the proposed IPPO-DM algorithm. \par

\begin{figure}
\centering
\includegraphics[width=0.9\linewidth]{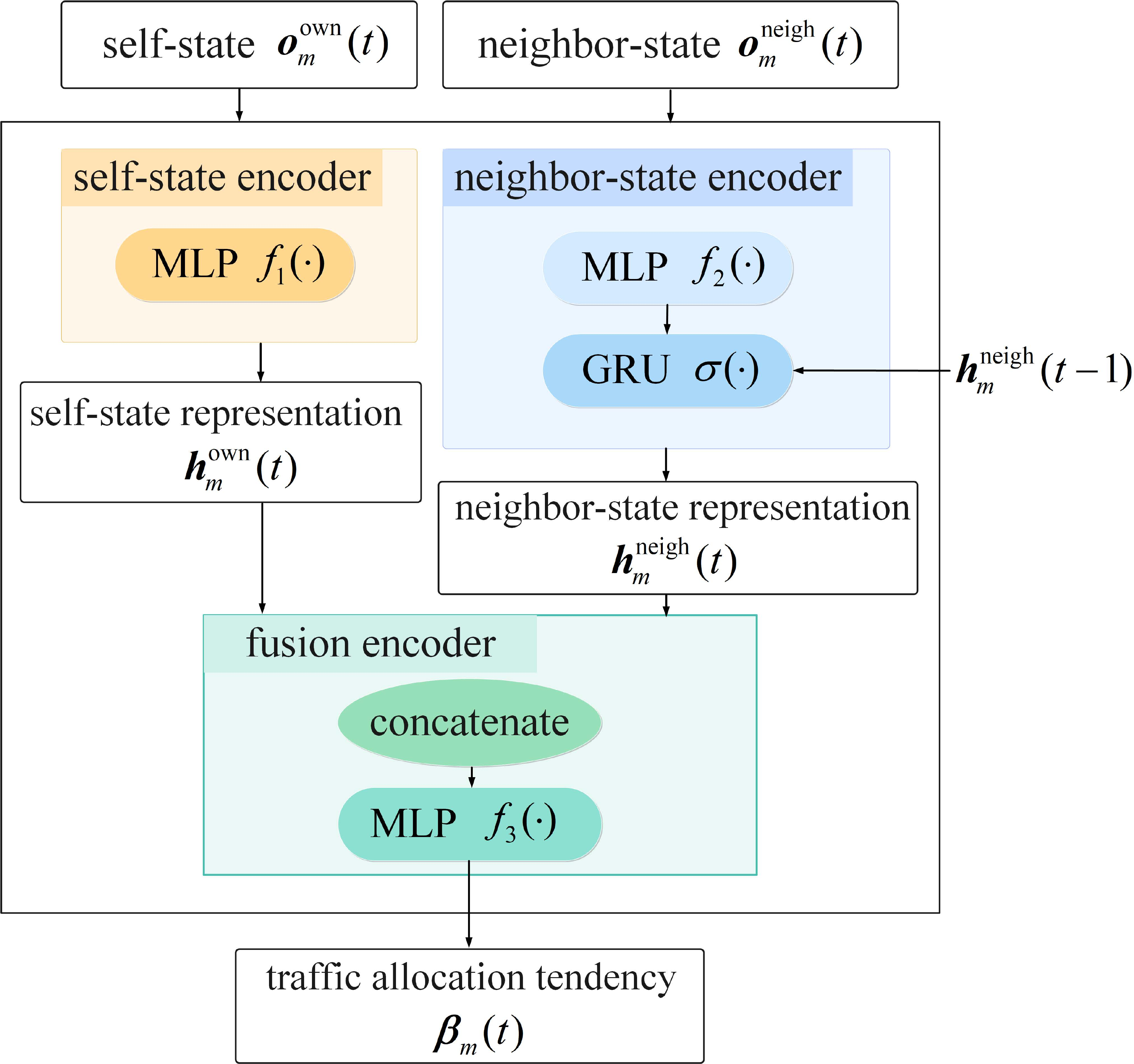}
\caption{Architecture of the Actor network.}
\label{actor}
\end{figure}

\emph{1) Actor–Critic Design:} Each UAV is required to complete traffic forwarding independently based on its local observations, without access to global state information. To address such distributed decision-making scenarios, we adopt IPPO as the policy optimization framework. IPPO is a lightweight and training-stable MARL method that enables decentralized policy learning.\par

In this setting, each UAV is modeled as an agent equipped with an Actor–Critic pair of networks. The Actor network is responsible for generating routing actions, whereas the Critic approximates the value function and provides value estimates to guide policy optimization. Since UAVs share highly consistent observation structures and decision logic, we employ a parameter-sharing mechanism, such that all UAVs utilize a common Actor and a common Critic network. This approach improves sample efficiency, enhances policy generalization, and reduces the overall training cost.

Within the IPPO framework, the Actor network of UAV $m$ is defined as $\mu({\bm o}_m(t)\mid\theta^\mu)$, where $\theta^\mu$ denotes the network parameters and ${\bm o}_m(t) = [{\bm o}_m^{\text{own}}(t), {\bm o}_m^{\text{neigh}}(t)]$ is the local observation. Given ${\bm o}_m(t)$ as input, the Actor outputs the traffic allocation tendency vector ${\bm \beta}_m(t)= [\beta_{m,0}(t), \ldots, \beta_{m,N}(t)]$, where $\beta_{m,0}(t)$ represents the preference for retaining traffic locally, and $\beta_{m,n}(t), \; n \in \{1,\ldots,N\}$ corresponds to the preferences for forwarding to the $n$-th neighbor.\par

In detail, the Actor network is designed with three encoding stages, as illustrated in Fig.~\ref{actor}. First, the self-state vector ${\bm o}_m^{\text{own}}(t)$ is processed by a Multi-Layer Perceptron (MLP) to obtain a high-dimensional self-state representation,
\begin{equation}
{\bm h}_m^{\text{own}}(t) = f_1({\bm o}_m^{\text{own}}(t)),
\label{self-state-encode}
\end{equation}where $f_1(\cdot)$ denotes the MLP. Meanwhile, the neighbor-state vector ${\bm o}_m^{\text{neigh}}(t)$ is first encoded by another MLP and subsequently passed through a Gated Recurrent Unit (GRU)~\cite{GRU} to capture temporal dynamics. This yields the neighbor-state representation,
\begin{equation}
{\bm h}_m^{\text{neigh}}(t) = \sigma \big(f_2({\bm o}_m^{\text{neigh}}(t)),  {\bm h}_m^{\text{neigh}}(t-1)\big),
\label{neighbor-state-encode}
\end{equation}where ${\bm h}_m^{\text{neigh}}(t-1)$ is the hidden state from the previous time slot, $f_2(\cdot)$ represents the MLP, and $\sigma(\cdot)$ denotes the GRU. Finally, the self-state and neighbor-state representations are concatenated and passed through a fusion MLP to extract joint features and generate the final traffic allocation tendency vector, which can be expressed as
\begin{equation}
{\bm \beta}_m(t) = f_3\big(\text{concat}({\bm h}_m^{\text{own}}(t) , {\bm h}_m^{\text{neigh}}(t))\big),
\label{fusion-encode}
\end{equation}where ${\rm concat}(\cdot)$ denotes the concatenation operation, and $f_3(\cdot)$ is the fusion MLP.\par

Furthermore, the Critic network is defined as $\nu({\bm o}_m(t)\mid\theta^\nu)$, where $\theta^\nu$ denotes the network parameters. The Critic network adopts the same encoding structure as the Actor to ensure consistency in feature representation. Unlike the Actor, it outputs a scalar value estimate $v_m(t) = \nu({\bm o}_m(t)\mid\theta^\nu),$ which reflects the expected long-term return of the current observation under the given policy.

\emph{2) Action Modeling Based on the Dirichlet Distribution:} In the defined action space, a routing decision involves allocating traffic among at most $N$ neighboring nodes or retaining it locally, subject to the strict constraint that allocation ratios must sum to one (i.e., the probability simplex). However, standard normalization methods like the Softmax function are insufficient for this task within the IPPO framework. While Softmax can ensure the output ratios sum to one, it yields a deterministic vector rather than a stochastic policy. Consequently, it fails to support the exploration mechanism and does not provide the probability density function required to compute the log-probabilities for policy optimization.\par

To construct a feasible stochastic policy, we employ the Dirichlet distribution. The Dirichlet distribution is a continuous probability distribution defined on the simplex, capable of generating vectors of non-negative proportions that naturally sum to one. Specifically, let the concentration parameters be ${\bm \alpha} = [\alpha_0, \ldots, \alpha_N]$ with $\alpha_n > 0,$ $\forall n \in \{0, \ldots, N\}$. The Dirichlet distribution parameterized by ${\bm \alpha}$ is written as ${\rm Dir}({\bm \alpha})$. If a vector ${\bm a} = [a_0, \ldots, a_N]$ is sampled from this distribution, denoted as ${\bm a} \sim {\rm Dir}({\bm \alpha})$, where each component satisfies $a_n > 0$ for all $n \in \{0, \ldots, N\}$ and $\sum_{n=0}^N a_n = 1$. The expectation and variance of each component $a_n$ are governed by the concentration parameters ${\bm \alpha}$, which are
\begin{align}
    &\mathbb{E}[a_n] = \frac{\alpha_n}{\hat{\alpha}}, \quad \forall n \in \{0, \ldots, N\}, \label{dirich-expec} \\
    &{\rm Var}[a_n] = \frac{\alpha_n(\hat{\alpha} - \alpha_n)}{\hat{\alpha}^2(\hat{\alpha} + 1)}, \quad \forall n \in \{0, \ldots, N\}, \label{dirich-var}
\end{align}where $\hat{\alpha} = \sum_{n=0}^N \alpha_n$. \par

In our design, the Dirichlet distribution is constructed from the Actor output ${{\bm \beta }_m}(t)$ to sample routing actions. To induce sparsity in traffic allocation, ${{\bm \beta }_m}(t)$ is first mapped into a sparse probability space using the Sparsemax function~\cite{sparsemax}, yielding
\begin{equation}
    {\bm \alpha }_m^{\text{ori}}(t) = \text{Sparsemax}({\bm \beta }_m(t)).
\end{equation}To further regulate exploration, a positive scaling factor $\rho$ is applied,
\begin{equation}
    {\bm \alpha }_m^{\text{scal}}(t) = \rho \cdot {\bm \alpha }_m^{\text{ori}}(t) + {\bm \alpha }_{\min}, 
    \label{alpha-scale}
\end{equation}where ${\bm \alpha }_{\min}$ is a constant vector with small positive entries to guarantee strict positivity. For cases where the number of available neighbors is smaller than $N$, invalid dimensions are masked by replacing them with a small constant $\epsilon$. The resulting concentration parameter vector is given by ${\bm \alpha }_m(t) = [\alpha_{m,0}(t), \ldots, \alpha_{m,N}(t)]$, with each component defined as
\begin{equation}
    \alpha_{m,n}(t) = 
    \begin{cases} 
        \alpha_{m,n}^{\text{scal}}(t), & 0 \leq n \leq |s_m(t)|, \\[6pt] 
        \epsilon, & \text{otherwise}, 
    \end{cases}
    \label{alpha}
\end{equation}and the final routing action is sampled as
\begin{equation}
    {\bm a}_m(t) \sim {\rm Dir}({\bm \alpha }_m(t)).
    \label{action-sample}
\end{equation}\par

Moreover, to enhance adaptability during execution, we introduce a traffic-aware action resampling mechanism. Specifically, the number of forwarding nodes is adaptively determined by the traffic scale. Given the queue length $q_m^{\text{sel}}(t)$, the number of forwarding nodes $\omega_m^{\text{sel}}(t)$ is computed as
\begin{equation}
    \omega_m^{\text{sel}}(t) = \min\left(\left\lceil \frac{q_m^{\text{sel}}(t)}{q_{\text{step}}} \right\rceil, N\right),
    \label{forward-nodes-number}
\end{equation}where $q_{\text{step}}$ is the step size in packets and $\lceil \cdot \rceil$ is the ceiling function. This ensures that the number of forwarding nodes increases proportionally with traffic load, up to $N$, thereby improving both efficiency and reliability.\par 

\begin{figure}
    \centering
    \includegraphics[width=0.9\linewidth]{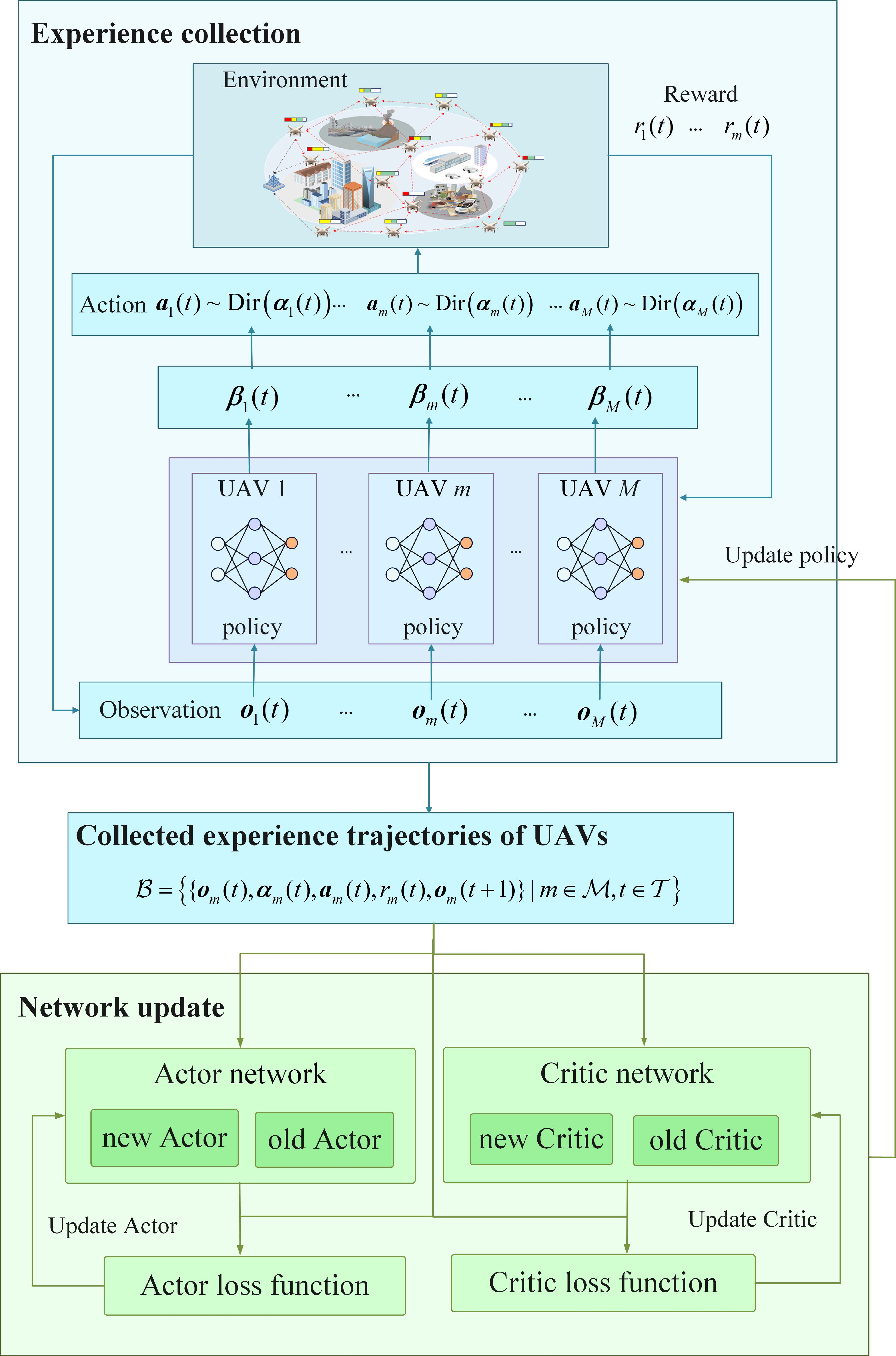}
    \caption{Experience Collection and Network Update.}
    \label{policy-update}
\end{figure}

Once the number of forwarding nodes $\omega_m^{\text{sel}}(t)$ has been determined, the next step is to select the corresponding forwarding paths. Specifically, from the action vector ${\bm a}_m(t)$, we exclude the first component $a_{m,0}(t)$ (representing the non-forwarding proportion) and retain the $\omega_m^{\text{sel}}(t)$ entries with the largest splitting ratios. The index set of these selected entries is denoted as
\begin{equation}
    {\cal I}_m^{\text{sel}}(t) = \text{TopI}\big({\bm a}_m(t) \setminus \{a_{m,0}(t)\}, \omega_m^{\text{sel}}(t)\big),
    \label{forward-nodes-index}
\end{equation}where $\text{TopI}(\cdot,\omega_m^{\text{sel}}(t))$ denotes the operation that extracts the indices of the $\omega_m^{\text{sel}}(t)$ largest elements in the input vector. Based on this, the filtered action vector $\tilde{\bm a}_m(t)$ is defined by keeping $a_{m,0}(t)$ and entries with indices in ${\cal I}_m^{\text{sel}}(t)$, while setting all others to zero. Finally, normalization produces the executable action vector,
\begin{equation}
    \bar a_{m,n}(t) = 
    \begin{cases} 
        \frac{\tilde a_{m,n}(t)}{\sum\limits_{n \in \{0\} \cup {\cal I}_m^{\text{sel}}(t)} \tilde a_{m,n}(t)}, & \text{if } n=0 \text{ or } n \in {\cal I}_m^{\text{sel}}(t), \\[10pt] 
        0, & \text{otherwise}. 
    \end{cases}
    \label{actual-action}
\end{equation}  

This resampling mechanism is applied only during the execution phase to enhance action efficiency. Although it modifies the original action structure, the resampled action retains the policy tendencies of the underlying distribution. Consequently, the rewards obtained still serve as faithful feedback to the original action distribution, ensuring that policy training is not misled by the resampling process.\par

\emph{3) Experience Collection and Policy Optimization:} As illustrated in Fig.~\ref{policy-update}, all UAV agents employ shared-parameter Actor networks and shared-parameter Critic networks. During training, experiences are gathered through repeated interactions between the agents and the environment, and the network parameters are periodically updated to maximize the long-term cumulative return.\par

In detail, a total of $N_{\rm epi}$ training episodes are conducted. In each episode, at each slot $t$, UAV $m$ feeds its local observation ${\bm o}_m(t)$ together with the previous encoded neighbor state ${\bm h}_m^{\text{neigh}}(t-1)$ into the Actor network, producing ${\bm \beta}_m(t)$. A Dirichlet distribution ${\rm Dir}({\bm \alpha}_m(t))$ is then constructed from ${\bm \beta}_m(t)$, and a routing action ${\bm a}_m(t)$ is sampled accordingly. The environment executes this action, returning the immediate reward $r_m(t)$ and the next observation ${\bm o}_m(t+1)$. After one episode, all interactions across agents and slots are aggregated into an experience buffer, ${\cal B} = \{ \{ {\bm o}_m(t), {\bm \alpha}_m(t), {\bm a}_m(t), r_m(t), {\bm o}_m(t+1) \} \mid m \in {\cal M}, t \in {\cal T}\}.$ At the same time, the Actor and Critic parameters used during trajectory collection are frozen and stored as the old networks. In the subsequent optimization phase, the updated Actor and Critic are referred to as the current networks, and their parameters are refined for $N_{\rm upd}$ update rounds following the IPPO algorithm.\par

For updating the Actor network, the objective is to maximize the expected return of the policy. The corresponding loss function is constructed based on three key elements, the log-probability of the sampled actions, the policy entropy, and the advantage. The log-probability of an action ${{\bm a}_m}(t)$ under the Dirichlet distribution ${\rm Dir}({\bm \alpha}_m(t))$ is given by
\begin{align}
    &\log p({{\bm a}_m}(t)|{{\bm \alpha} _m}(t)) 
    = \log \Gamma ( {\sum_{n = 0}^N \alpha_{m,n}(t)} )- \nonumber \\
    &\quad \sum_{n = 0}^N \log \Gamma (\alpha_{m,n}(t)) + \sum_{n = 0}^N (\alpha_{m,n}(t) - 1)\log a_{m,n}(t),
    \label{log-prob-action}
\end{align}where $\Gamma(\cdot)$ is the Gamma function. The entropy of ${\rm Dir}({\bm \alpha}_m(t))$ is expressed as
\begin{align}
    H({{\bm a}_m}(t)) 
    = &- \sum_{n = 0}^N \big(\alpha_{m,n}(t) - 1\big) 
       \Big[ \psi(\alpha_{m,n}(t)) - \psi(\hat \alpha_m(t)) \Big] \nonumber \\
    &\quad - \ln C({\bm \alpha}_m(t)),
    \label{entropy-action}
\end{align}where $\hat \alpha_m(t) = \sum_{n=0}^N \alpha_{m,n}(t)$ represents the sum of concentration parameters, $C({\bm \alpha}_m(t)) = \Gamma(\hat \alpha_m(t))/\prod_{n=0}^N \Gamma(\alpha_{m,n}(t))$ is the normalization constant corresponding to the multivariate extension of the Beta function, and $\psi(\cdot)$ denotes the Digamma function.\par

The advantage function is estimated using Generalized Advantage Estimation (GAE)~\cite{GAE}. The temporal-difference (TD) error at slot $t$ is defined as
\begin{equation}
    \delta_m(t) = r_m(t) + \gamma v_m^{\rm old}(t+1) - v_m^{\rm old}(t),
    \label{TD-error}
\end{equation}where $0 \leq \gamma \leq 1$ is the discount factor, and $v_m^{\rm old}(\cdot)$ denotes the value estimate provided by the old Critic network. The corresponding advantage is recursively computed as
\begin{equation}
    A_m(t) = \delta_m(t) + \gamma \lambda A_m(t+1),
    \label{advantage}
\end{equation}with $0 \leq \lambda \leq 1$ controlling the GAE smoothing.\par

Based on the above definitions, the policy ratio between the current and old Actor networks for the same action is computed as
\begin{align}
    \varpi_m(t) 
    = \exp \Big(&\log p({\bm a}_m(t)|{\bm \alpha}_m(t)) \nonumber \\
     & - \log p({\bm a}_m(t)|{\bm \alpha}_m^{\rm old}(t))\Big),
    \label{new-old-policy-ratio}
\end{align}where $\exp(\cdot)$ denotes the exponential function and ${\bm a}_m(t)$ is the action stored in the experience buffer. To constrain overly large updates, clipping is applied,
\begin{align}
    L_m^{\rm clip}(t) 
    = \min \big\{& \varpi_m(t) A_m(t), \nonumber \\
    &{\rm clip}(\varpi_m(t), 1-\varepsilon_1, 1+\varepsilon_1) A_m(t) \big\},
    \label{actor-loss-clip}
\end{align}where ${\rm{clip}}(\cdot)$ denotes the clipping function and ${\varepsilon _1}$ is the clipping threshold. To encourage sufficient exploration, an entropy regularization term is added, yielding the final Actor loss
\begin{equation}
    L^{\rm actor} 
    = - \frac{1}{TM} \sum_{t=1}^T \sum_{m=1}^M 
      \Big( L_m^{\rm clip}(t) - \wp H({\bm a}_m(t)) \Big),
    \label{actor-loss}
\end{equation}where $\wp$ controls the weight of entropy regularization in the policy update.

\begin{algorithm}[t]
\label{alg1}
\caption{Training procedure of the IPPO-DM algorithm}
Initialize $\theta^{\mu}$ of the actor network $\mu({\bm o}_m(t)\mid\theta^{\mu})$.\;
Initialize $\theta^{v}$ of the critic network $v({\bm o}_m(t)\mid\theta^{v})$.\;
Initialize the experience buffer $\mathcal{B}$.\;
\ForE{$i= \{1,2,\ldots, N_{\rm epi}\}$}{
  \ForE{$t = \{1,2,\ldots,T\}$}{
    Obtain local observations $\{{\bm o}_m(t)\mid m\in\mathcal{M}\}$.\;
    \ForE{$m = \{1,2,\ldots,M\}$}{
      Compute $\bm{\beta}_m(t) \gets \mu({\bm o}_m(t)\mid\theta^{\mu})$.\;
      Construct the Dirichlet distribution $\mathrm{Dir}(\bm{\alpha}_m(t))$ based on $\bm{\beta}_m(t)$.\;
      Sample and execute action ${\bm a}_m(t) \sim \mathrm{Dir}(\bm{\alpha}_m(t))$.\;
      Receive reward $r_m(t)$.\;
      Store $\{{\bm o}_m(t),\bm{\alpha}_m(t),{\bm a}_m(t),r_m(t),{\bm o}_m(t+1)\}$ in $\mathcal{B}$.\;
    }
  }
  Set $\theta^{\mu}_{\text{old}} \gets \theta^{\mu}$, $\theta^{v}_{\text{old}} \gets \theta^{v}$.\;
  \ForE{$n = \{ 1,2,\ldots,N_{\rm epi}\}$}{
    Compute $L^{\text{actor}}$ according to Eq.~\eqref{actor-loss}.\; update $\theta^{\mu}$.\;
    Compute $L^{\text{critic}}$ according to Eq.~\eqref{critic-loss}.\; update $\theta^{v}$.\;
  }
}
\end{algorithm}

For updating the Critic network, a value clipping strategy is adopted to stabilize training and prevent oscillations~\cite{PPO}. The clipped value estimate is defined as
\begin{equation}
    v_m^{\rm clip}(t) = v_m^{\rm old}(t) + {\rm clip}\big(v_m(t) - v_m^{\rm old}(t), -\varepsilon_2, \varepsilon_2\big),
    \label{clip-value}
\end{equation}where $v_m^{\rm old}(t)$ is the value estimate from the old Critic network, $v_m(t)$ is the estimate from the current Critic network, and $\varepsilon_2$ is the clipping threshold. The Critic loss is then defined as
\begin{align}
    L^{\rm critic} = - \frac{1}{TM}\sum_{t=1}^T \sum_{m=1}^M 
    \min \Big\{ &\big(v_m(t) - \hat v_m(t)\big)^2, \nonumber \\
    &\big(v_m^{\rm clip}(t) - \hat v_m(t)\big)^2 \Big\},
    \label{critic-loss}
\end{align}where the target value is $\hat v_m(t) = r_m(t) + \gamma v_m^{\rm old}(t+1)$. By taking the minimum of the original and clipped squared errors, the update magnitude of the Critic is constrained, reducing the risk of value overestimation and gradient instability.  

IPPO-DM follows the centralized training and decentralized execution paradigm, and the overall training procedure is outlined in Algorithm~\ref{alg1}.

\section{Numeric Results}
This section presents simulation-based comparisons between the proposed IPPO-DM algorithm and benchmark algorithms under varying network loads, different UAV network scale, and different constraints on the number of forwarding nodes. The evaluation focuses on the algorithm’s ability to jointly ensure on-time packet delivery and minimize packet loss, thereby providing a comprehensive assessment of its reliability and stability in complex network environments.

\textit{1) System settings:} In this work, we consider a three-dimensional environment of size $1200 \, \text{m} \times 1200 \, \text{m} \times 200 \, \text{m}$. UAVs are distributed in the airspace above $100 \, \text{m}$ altitude. The GBS is fixed at the edge of the ground environment with an altitude of $0 \, \text{m}$. \par 

Throughout the UAV flights, the three-dimensional Gauss-Markov mobility model~\cite{UAV-mobile} is adopted to characterize the UAV trajectories, where the UAV velocity is constrained within a maximum of $[50, 50, 20]\ {\rm  m/s}$ and a minimum of $[15, 15, 5]\ {\rm  m/s}$. The average velocity $[25, 25, 10]\ {\rm  m/s}$ is further adopted as the reference velocity in the mobility model to stabilize the trajectory evolution. In each time slot, a UAV initiates a sensing or inspection task with probability ${p^{{\rm flow}}}$, which generates traffic that needs to be transmitted to the GBS. The task traffic size is randomly selected within the range $[0.5, 2] \, \text{MB}$. Correspondingly, the deadline of each task is set to be linearly proportional to its data size, ranging from $1.5 \, \text{s}$ to $3 \, \text{s}$. In addition, the maximum buffer size of each UAV's communication queue is randomly assigned within the range $[3000, 5000]$ packets. Other key environment parameters are summarized in Table~\ref{table2}.   \par 

For the training setup of the IPPO-DM algorithm, we implement the simulations using PyTorch~\cite{pytorch}, and employ the AdamW optimizer~\cite{adamw} to update the parameters of both the Actor and Critic networks. The learning rate and weight decay coefficient are set to $2 \times 10^{-4}$ and $1 \times10^{-3}$, respectively. Other training-related parameters are listed in Table~\ref{table3}.   \par

\begin{table}[!t]
\renewcommand{\arraystretch}{1.4}
\caption{Environment Parameters}
\label{table2}
\centering
\begin{tabular}{|m{5.4cm}|>{\centering\arraybackslash}m{2.5cm}|}
\hline
\textbf{Parameter} & \textbf{Value} \\ \hline
Number of UAV nodes & $M=35$ \\ \hline
Duration of time slot & $\Delta t=0.05\ \rm s$ \\ \hline
Duration of flight time & $T=9\ \rm s$ \\ \hline
Task arrival probability & $p^{\rm flow}=0.03$ \\ \hline
Per-packet payload & $I_{\rm pkt}=1500\ \rm Bytes$ \\ \hline
Channel gain at the reference distance & $\beta_0=-50\ \rm dB$ \\ \hline
Maximum number of UAV neighbors & $N=8$ \\ \hline
Maximum transmit power of UAVs & $P_m^{\max }=30\ \rm dBm$ \\ \hline
White noise power spectral density & $N_0=-174\ \rm dBm/Hz$ \\ \hline
Number of available orthogonal sub-channels & $B=25$ \\ \hline
Bandwidth of each sub-channel & $b=20\ \rm MHz$ \\ \hline
Channel S-curve parameters & 
\makecell{$D_1=9.61,$ \\ $D_2=0.15$} \\ \hline
Path loss exponent & $h=2$ \\ \hline
Excessive pathloss exponent & 
\makecell{$\eta_{\rm LoS}=1\ \rm dB,$ \\ $\eta_{\rm NLoS}=20\ \rm dB$} \\ \hline
Carrier frequency & $f_c=2.4\ \rm GHz$ \\ \hline
Minimum required SINR threshold & $g^{\min}=11\ \rm dB$ \\ \hline
\end{tabular}
\end{table}

\begin{table}[!t]
\renewcommand{\arraystretch}{1.4}
\caption{IPPO-DM Algorithm Parameters}
\label{table3}
\centering
\begin{tabular}{|m{5.4cm}|>{\centering\arraybackslash}m{2.5cm}|}
\hline
\textbf{Parameter} & \textbf{Value} \\ \hline
Output feature dimension of MLP $f_1(\cdot)$ & $128$ \\ \hline
Output feature dimension of MLP $f_2(\cdot)$ & $256$ \\ \hline
Output feature dimension of GRU $\sigma(\cdot)$ & $256$ \\ \hline
\multicolumn{1}{|m{5.5cm}|}{Minimum/Maximum reward weight parameter} & \makecell{$w_{\min}=0.2,$ \\ $w_{\max}=0.8$}  \\ \hline
\multicolumn{1}{|m{5.5cm}|}{Minimum/Maximum buffer size of the highest-priority sub-queue} & 
\makecell{$q_{\min}=1,$ \\ $q_{\max}=1000$} \\ \hline
\multicolumn{1}{|m{5.5cm}|}{Penalty factor associated with data-flow priority} & \makecell{$-r^{\rm st}(1)=-0.5,$ \\ $-r^{\rm st}(2)=-0.3,$ \\ $-r^{\rm st}(3)=-0.05$}\\ \hline
Fixed penalty parameter for non-forwarding & $-r_0^{\rm tol}=-0.1$ \\ \hline
Reward curve slope control parameters & \makecell{$k_1=3.5,$ \\ $k_2= -0.35$}\\ \hline
Scaling factor & \makecell{$\rho=30,$ \\ $\alpha_{\min}= 0.5$} \\ \hline
Small positive value & $\epsilon=1\times 10^{-8}$ \\ \hline
Packet step size & $q_{\rm step}=300$ \\ \hline
Total training episodes  & $N_{\rm epi}=3500$ \\ \hline
Network parameter update rounds & $N_{\rm upd}=5$ \\ \hline
Discount factor  & $\gamma=0.95$ \\ \hline
GAE smoothing coefficient  & $\lambda=0.95$ \\ \hline
Clipping coefficient & \makecell{$\epsilon_1=0.05,$ \\ $\epsilon_2= 0.2$} \\ \hline
Entropy coefficient & $\varphi=0.01$ \\ \hline
\end{tabular}
\end{table}

\textit{2) Benchmark Algorithms: } In the simulations, two representative routing algorithms are selected as baselines. All experiments are independently executed $50$ times under identical environments, and the average results are reported. The baseline algorithms are described as follows:  

\begin{itemize}
    \item \textbf{Heuristic Traffic Splitting Algorithm}: This algorithm employs a static traffic splitting strategy. Packets are uniformly distributed among all candidate neighbors that are closer to the GBS than the current node. This approach aims to balance the forwarding load while ensuring geographic progress toward the destination.  
    
    \item \textbf{Greedy Shortest-Path Algorithm}: This algorithm adopts a deterministic single-path strategy. At each step, packets from the non-empty highest-priority sub-queue are forwarded exclusively to the single neighbor closest to the GBS. This baseline represents the strategy prioritizing transmission delay minimization. 

\end{itemize}  

It is worth noting that, to ensure fairness in comparison, we apply a unified traffic-aware action resampling mechanism (eq.~\eqref{forward-nodes-number}–eq.~\eqref{actual-action}) during the execution phase to both the proposed IPPO-DM and the heuristic traffic splitting algorithm. \par

\begin{figure}
    \centering
    \includegraphics[width=0.85\linewidth]{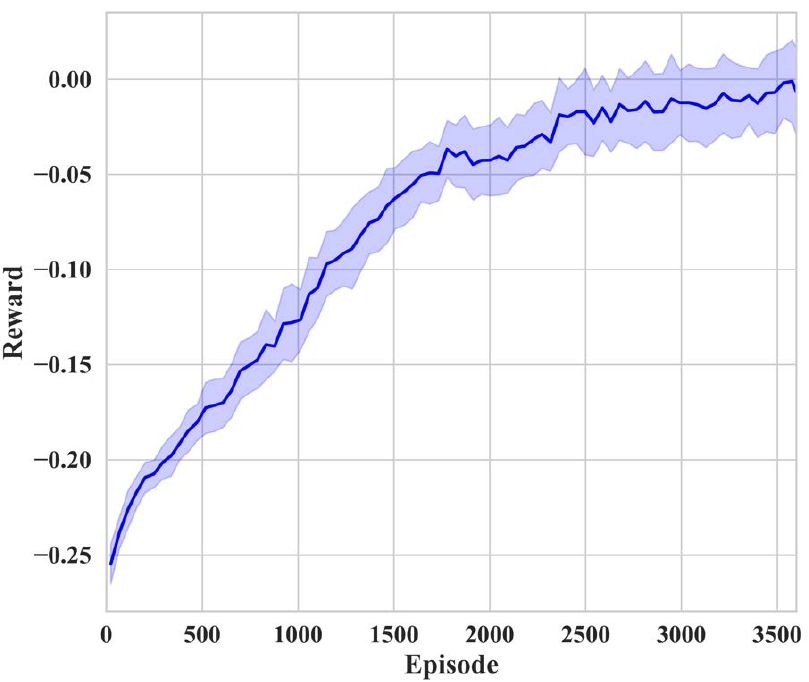}
    \caption{Convergence of the IPPO-DM algorithm.}
    \label{convergence}
\end{figure}

\begin{figure}
    \centering
    \includegraphics[width=0.85\linewidth]{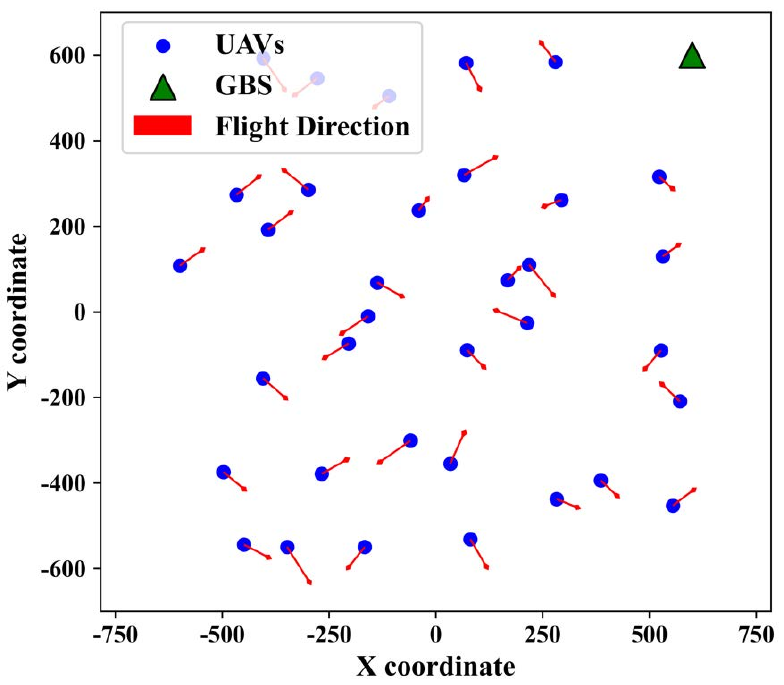}
    \caption{Spatial Distribution of GBS and UAV Trajectories.}
    \label{GBS-UAV-distri}
\end{figure}

\subsection{Performance Evaluation}

Fig.~\ref{convergence} illustrates the convergence of the proposed IPPO-DM algorithm. As the number of training episodes increases, the reward gradually rises and eventually stabilizes, indicating that the algorithm is capable of achieving stable policy learning throughout the training process.  \par

Fig.~\ref{GBS-UAV-distri} depicts the position of the GBS in the simulation environment, together with the spatial distribution and flight directions of the UAVs. It can be observed that the UAVs are randomly distributed within the simulation environment and move independently in different directions, thereby emulating the dynamic characteristics of UAVs. \par

\begin{figure*}[!t]
\centering
\subfloat[Cumulative packet arrival ratio]{
    \includegraphics[width=0.39\textwidth]{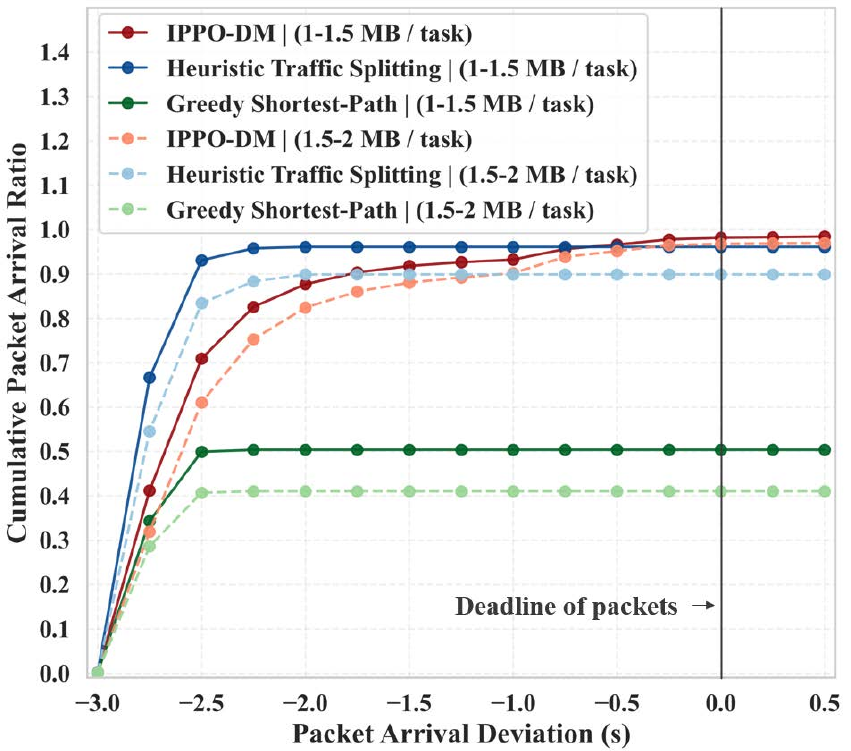}
    \label{TaskFeaPackDistri}
}
\hfil
\subfloat[Cumulative task arrival ratio]{
    \includegraphics[width=0.39\textwidth]{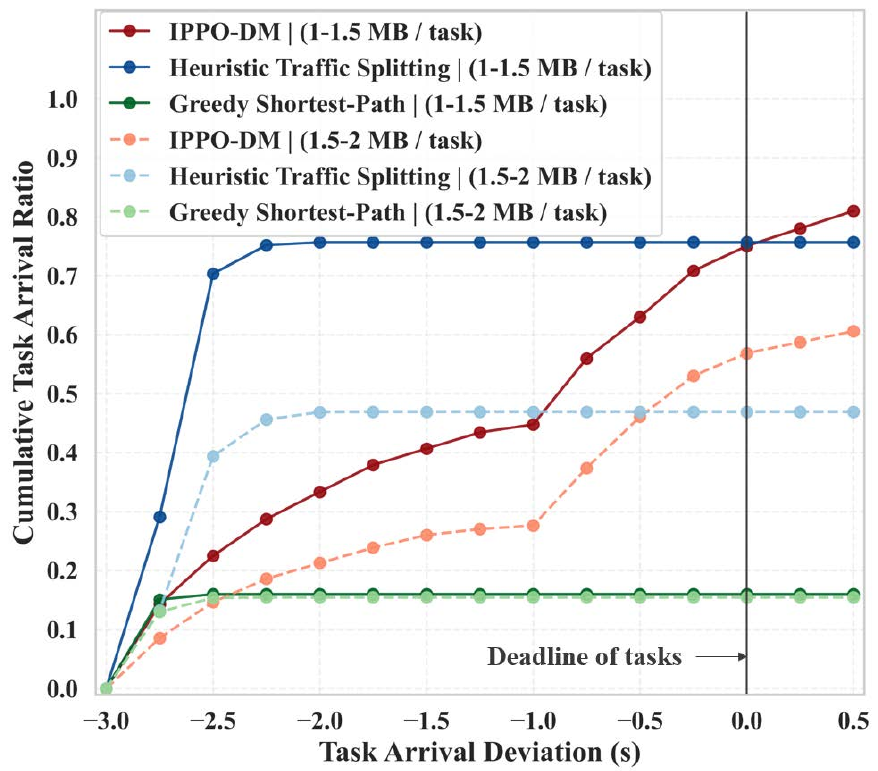}
    \label{TaskFeaTaskDistri}
}

\vspace{0cm} 

\subfloat[On-time packet delivery ratio]{
    \includegraphics[width=0.42\textwidth]{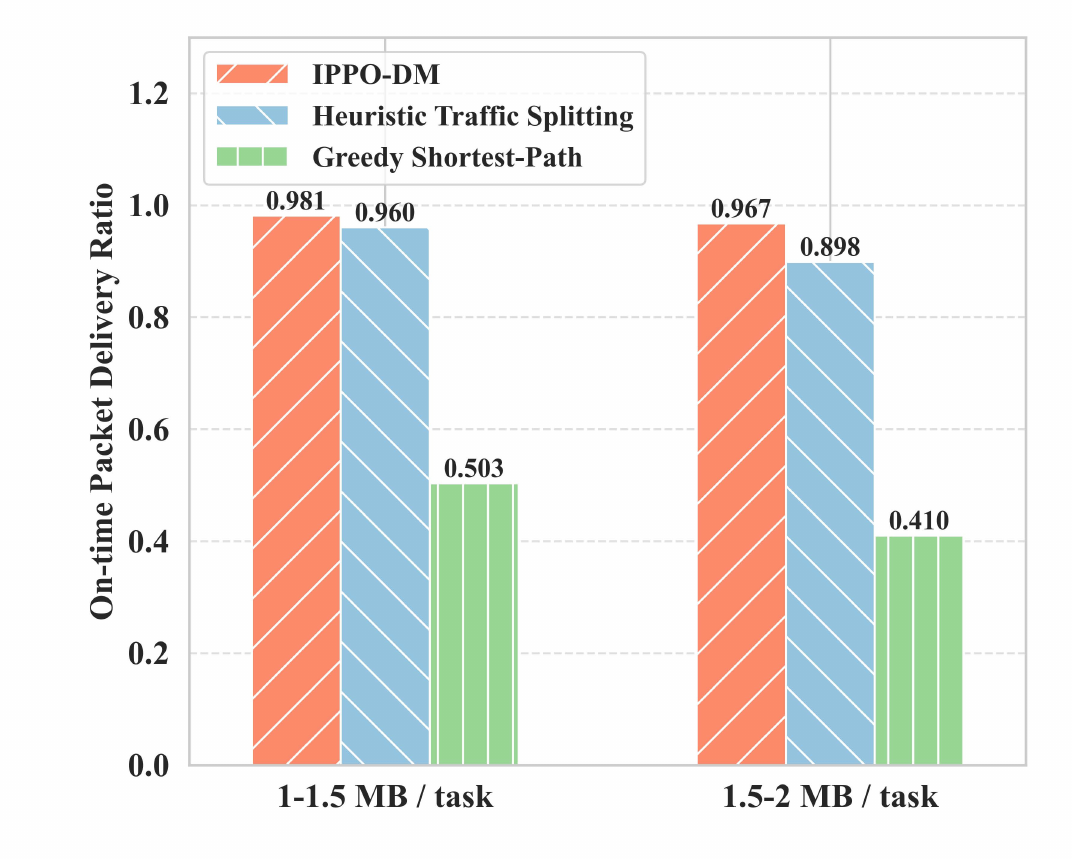}
    \label{TaskFeaPDR}
}
\hfil
\subfloat[Packet loss ratio]{
    \includegraphics[width=0.42\textwidth]{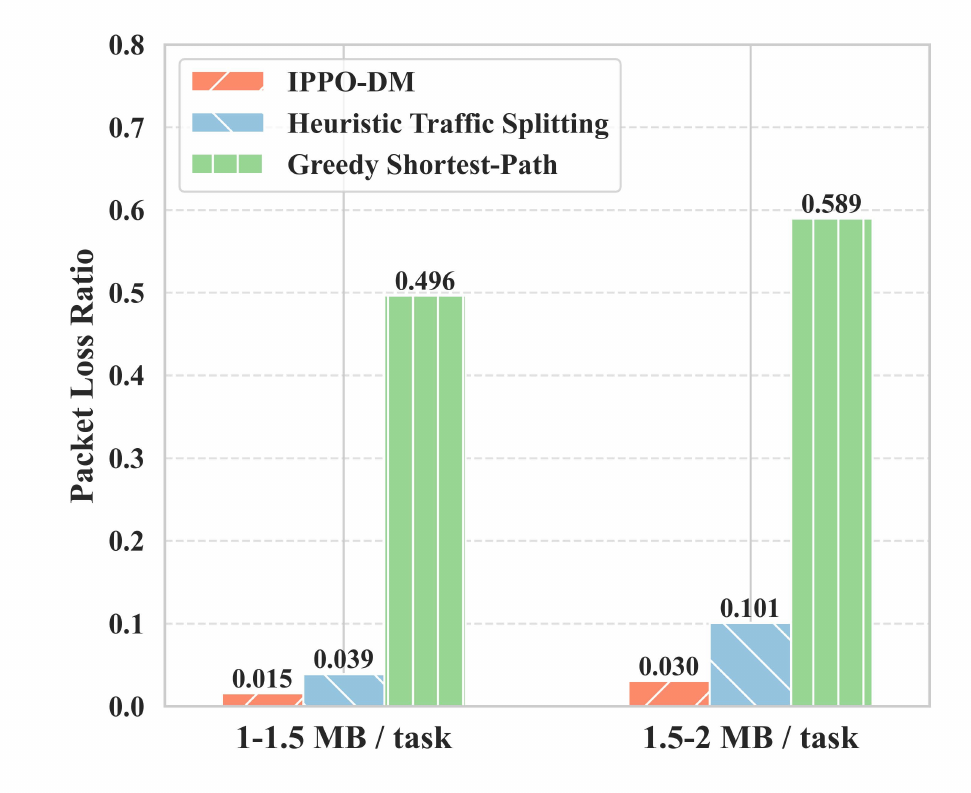}
    \label{TaskFeaDR}
}

\caption{Performance comparison under varying load conditions.}
\label{TaskFeaAll}
\end{figure*}

Fig.~\ref{TaskFeaAll} compares the performance of the three algorithms under low-load ($[1–1.5]$ MB/task) and high-load ($[1.5–2]$ MB/task) conditions. Specifically, Figs.~\ref{TaskFeaAll}\subref{TaskFeaPackDistri} and \subref{TaskFeaTaskDistri} depict the cumulative on-time arrival ratios at the packet and task levels, respectively, while Figs.~\ref{TaskFeaAll}\subref{TaskFeaPDR} and \subref{TaskFeaDR} present the overall on-time packet delivery and loss ratios.\par

 Fig.~\ref{TaskFeaAll}\subref{TaskFeaPackDistri} shows the cumulative packet arrival ratio, and Fig.~\ref{TaskFeaAll}\subref{TaskFeaTaskDistri} shows the cumulative task arrival ratio. The x-axis denotes the deviation between actual arrival time and deadline, where negative values indicate early arrivals and positive values indicate delayed arrivals. The y-axis represents the cumulative proportion of successful arrivals within a given deviation. All algorithms show performance degradation as the load increases. The heuristic traffic splitting algorithm performs well initially but saturates early, as its rigid splitting strategy causes packet loss on congested links. In contrast, IPPO-DM maintains a robust rising trajectory near the deadline, eventually stabilizing at a higher level. This superiority stems from its adaptive traffic splitting strategy, which effectively balances link utilization and mitigates congestion-induced loss. The greedy shortest-path algorithm performs the worst due to single-path congestion. Notably, in the task-level cumulative arrival curve (Fig.~\ref{TaskFeaAll}\subref{TaskFeaTaskDistri}), the IPPO-DM curve exhibits a sharp turning point at $x=-1$ (i.e., one second before the deadline), where the slope increases substantially. This validates the effectiveness of the priority mechanism (Eq.~\eqref{traffic_prio}), which successfully accelerates urgent tasks towards the GBS as they approach their deadlines. These trends are further corroborated by the overall statistics in Figs.~\ref{TaskFeaAll}\subref{TaskFeaPDR} and \subref{TaskFeaDR}. IPPO-DM consistently achieves the highest on-time delivery ratio and the lowest packet loss ratio across both load scenarios. Even under heavy load, it significantly outperforms the baselines, demonstrating that its traffic-aware adaptive splitting capability ensures both transmission reliability and delay satisfaction.\par

\begin{figure}
    \centering
    \includegraphics[width=0.95\linewidth]{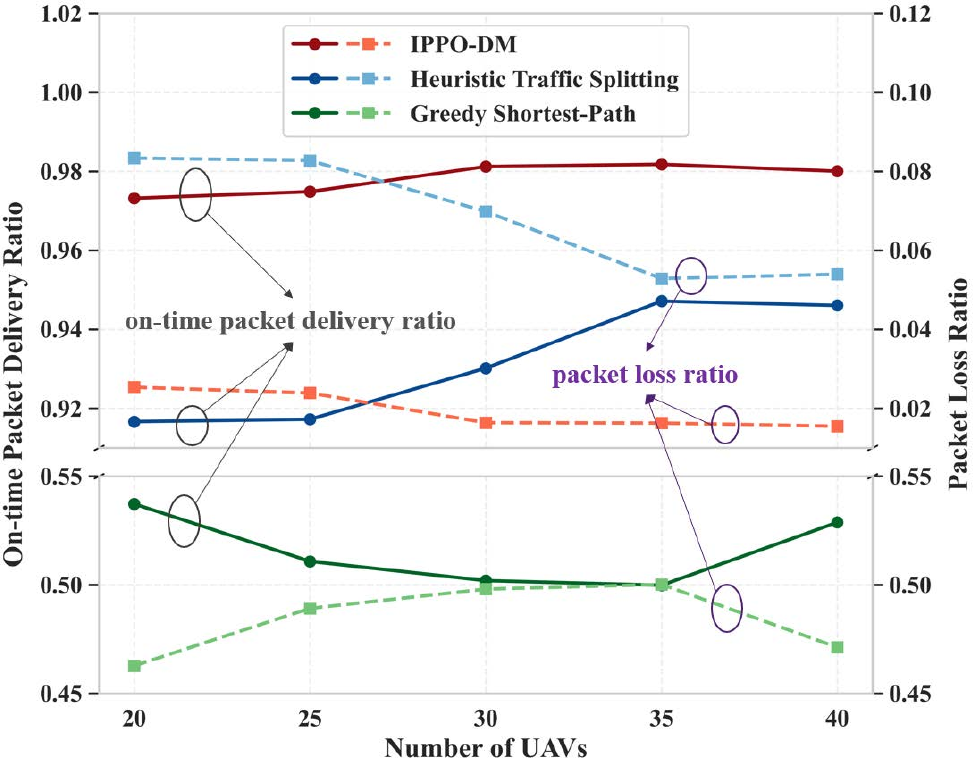}
    \caption{Performance under different numbers of UAVs.}
    \label{uavNumPDRDR}
\end{figure}

\begin{figure}
    \centering
    \includegraphics[width=0.95\linewidth]{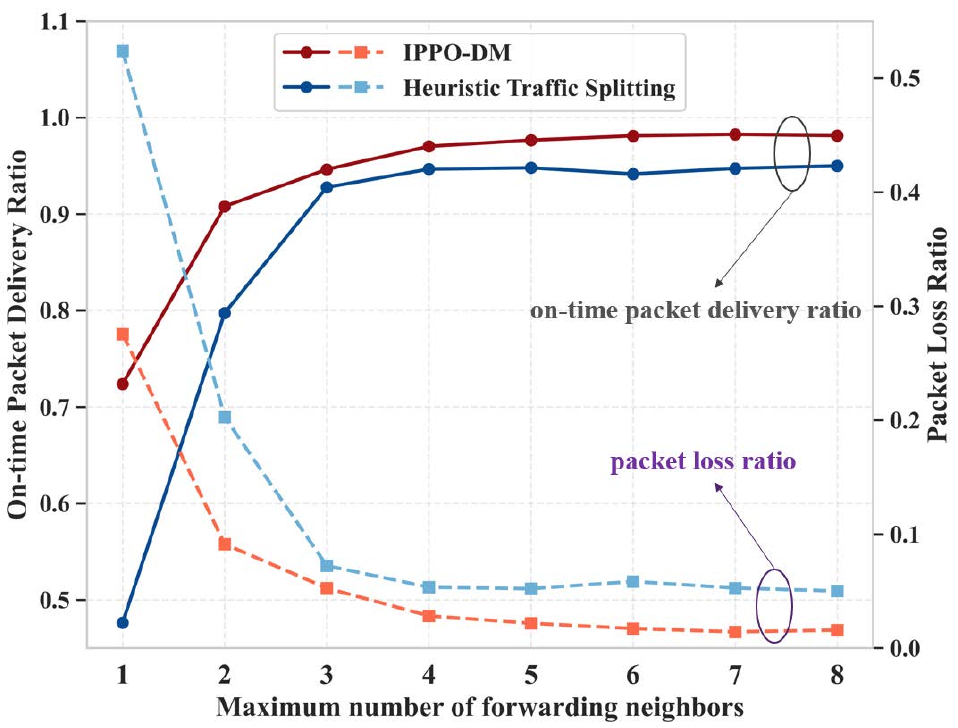}
    \caption{Performance under different numbers of candidate forwarding nodes.}
    \label{transNodePDRDR}
\end{figure}

Fig.~\ref{uavNumPDRDR} compares the performance of the three algorithms under varying network scales. As the number of UAVs increases, both IPPO-DM and the heuristic traffic splitting algorithm exhibit significant improvements in on-time packet arrival ratios and reductions in packet loss, eventually reaching a performance plateau. This trend stems from increased routing diversity: a denser network provides more forwarding options, enabling better load balancing and congestion alleviation. Crucially, IPPO-DM consistently maintains a lower packet loss ratio than the heuristic traffic splitting algorithm. This advantage arises from IPPO-DM’s active perception of queue states, which prevents traffic allocation to potentially congested links, thereby ensuring stability. In contrast, the greedy shortest-path algorithm fails to benefit from the increased network scale due to the rigidity of its single-path strategy, limiting its resource utilization.\par

Fig.~\ref{transNodePDRDR} illustrates the impact of the maximum number of forwarding candidates ($N$) on the two traffic splitting algorithms. It is observed that as $N$ increases, both algorithms show progressive improvements in on-time arrival ratios and reductions in packet loss, stabilizing once $N$ reaches five. When the candidate set is small, the limited decision space restricts path selection flexibility, easily leading to local congestion. Expanding the candidate set grants algorithms greater freedom in traffic allocation, significantly alleviating link bottlenecks. However, beyond the threshold of $N=5$, the marginal benefit of additional neighbors diminishes, indicating that a moderate number of next-hop candidates is sufficient to achieve great load balancing.\par

\section{Conclusions}

In this paper, a novel traffic-adaptive multipath routing framework was proposed, which enables each UAV to dynamically split and forward traffic flows across multiple next-hop neighbors. An on-time packet delivery ratio maximization problem was formulated to optimize the traffic splitting ratios. To solve this problem, the IPPO-DM algorithm was developed, which leverages MARL to achieve adaptive traffic splitting. Simulation results demonstrated that, across diverse network loads, UAV scales, and forwarding constraints, the proposed framework and algorithm outperformed benchmark schemes. Specifically, significant reductions in packet loss were compared to single-path routing due to effective congestion mitigation. Furthermore, the traffic-adaptive multipath scheme can stringently guarantee latency requirements by the proposed IPPO-DM. For future work, the integration of spatial-temporal topology prediction mechanisms will be investigated to further enhance routing stability in highly dynamic environments.


{
\bibliographystyle{IEEEtran}
\bibliography{IEEEabrv,bibe}
}

\end{document}